\documentclass[acmsmall]{acmart}
\renewcommand\footnotetextcopyrightpermission[1]{} 
\acmYear{00}
\acmMonth{0}

\usepackage{graphicx}
\usepackage{amsthm}

\newcounter{subsubsubsection}[subsubsection]
\renewcommand\thesubsubsubsection{\thesubsubsection.\arabic{subsubsubsection}}

\newcommand{\subsubsubsection}[1]{%
  \refstepcounter{subsubsubsection}%
  \paragraph{\thesubsubsubsection\quad #1}%
}
\AtBeginDocument{%
  }
\usepackage{amsmath}
\usepackage{enumitem}
\usepackage{tikz}
\usepackage{multirow}
\usepackage{soul}
\usepackage{array} 

\begin{document}
\title{Computationally Efficient Diffusion Models in Medical Imaging: A Comprehensive Review}
\author{Abdullah}
\email{abdullah@my.jcu.edu.au}
\orcid{0000-0001-7515-5371}
\affiliation{%
  \institution{James Cook University}
  \city{Cairns}
  \state{Queensland}
  \country{Australia}
}
\author{Tao Huang}
\email{tao.huang1@jcu.edu.au}
\orcid{0000-0002-8098-8906}
\affiliation{%
  \institution{James Cook University}
  \city{Cairns}
  \state{Queensland}
  \country{Australia}
}

\author{Ickjai Lee}
\email{ickjai.lee@jcu.edu.au}
\orcid{0000-0002-6886-6201}
\affiliation{%
  \institution{James Cook University}
  \city{Cairns}
  \state{Queensland}
  \country{Australia }
}
\author{Euijoon Ahn}
\email{euijoon.ahn@jcu.edu.au}
\orcid{https://orcid.org/0000-0001-7027-067X}
\authornotemark[1]
\affiliation{%
  \institution{James Cook University}
  \city{Cairns}
  \state{Queensland}
  \country{Australia}
}

\begin{abstract}

The diffusion model has recently emerged as a potent approach in computer vision, demonstrating remarkable performances in the field of generative artificial intelligence. 
%
Capable of producing high-quality synthetic images, diffusion models have been successfully applied across a range of applications. 
%
However, a significant challenge remains with the high computational cost associated with training and generating these models.
%
%
%
This study focuses on the efficiency and inference time of diffusion-based generative models, highlighting their applications in both natural and medical imaging. We present the most recent advances in diffusion models by categorizing them into three key models: the Denoising Diffusion Probabilistic Model (DDPM), the Latent Diffusion Model (LDM), and the Wavelet Diffusion Model (WDM). These models play a crucial role in medical imaging, where producing fast, reliable, and high-quality medical images is essential for accurate analysis of abnormalities and disease diagnosis. 
%
We first investigate the general framework of DDPM, LDM, and WDM and discuss the computational complexity gap filled by these models in natural and medical imaging. 
We then discuss the current limitations of these models as well as the opportunities and future research directions in medical imaging.

\end{abstract}
\ccsdesc[300]{Computing methodologies~Artificial intelligence}
\ccsdesc[300]{Computing methodologies~Computer vision}
\ccsdesc[300]{Applied computing~Life and medical science}
\ccsdesc[300]{Applied computing~Computational biology}
\ccsdesc[300]{Applied computing~Imaging}
\keywords{Generative Model, Diffusion model, Wavelet Diffusion Model, Medical Imaging}
\maketitle
\section{Introduction}
Deep generative models have demonstrated remarkable capabilities in generating new and realistic images \cite{RN187}. These models include Generative Adversarial Networks (GANs) \cite{RN187}, Variational Autoencoder Encoder (VAE) \cite{RN189}, and Energy-Based Models (EBMs) \cite{RN191}. GANs have garnered great attention among these models due to their training process involving a dual neural network system: a generator and a discriminator. The generation network produces and synthesizes data from random noise, while a discrimination network evaluates the authenticity of the data by differentiating between real and generated synthetic data. These networks work simultaneously to produce realistic synthetic data and are used for multiple tasks such as image generation \cite{A21}, image classification \cite{A22}, image segmentation \cite{A24}, and object detection \cite{A23}. However, GANs are still limited in producing high-quality synthetic data when the generator and discriminator fail to attain a stable equilibrium during training. It hinders the accurate generation of realistic data. GANs can also lead to model collapse, where the generator struggles to produce diverse outputs, resulting in repetitive or similar samples. The model collapses when the generator dominates the discriminator network, receiving insufficient feedback to guide its learning effectively \cite{RN196}. Moreover, GANs are limited in parallel processing due to dependencies in their alternating training process between the generator and discriminator. 

Recently, diffusion-based generative models \cite{RN192, RN193, RN195, RN217} have been examined as State-Of-The-Art (SOTA) in generating synthetic data due to their stable training performance and their superior performance in generating high-quality synthetic data compared to GANs. Diffusion models have been successfully applied in numerous applications, such as image synthesis \cite{RN228, RN199, RN224}, video generation \cite{RN204}, Natural Language Processing (NLP) \cite{RN205}, text generation \cite{RN206}, image denoising \cite{RN224, RN208},  image inpainting \cite{RN210}, image super-resolution \cite{RN226, RN209}, image segmentation~\cite{RN211, RN212}, image-to-image translation \cite{RN321}, and classification \cite{RN270, RN213}. Diffusion models effectively address the prominent limitations of GANs. These models progressively add Gaussian noise to the input image over a series of different time steps until the image becomes completely noisy. The model then learns to reconstruct the original input image from this fully noisy state.

More recently, Denoising Diffusion Probabilistic Models (DDPMs) have been introduced, surpassing the performance of traditional diffusion models \cite{RN224}. DDPMs incorporate a denoising technique to enhance the modeling of complex data distributions, improving sample quality and training stability compared to earlier diffusion models. This enhancement enables DDPMs to achieve superior performance for various generative modeling tasks. DDPMs have been successfully applied to many different imaging analysis tasks such as image generation \cite{RN294, RN259, RN238}, image segmentation \cite{RN401, RN426, RN417, RN418}, image inpainting \cite{RN265, RN266}, image classification \cite{RN270, RN477, RN427}, image translation \cite{RN273, RN448, RN327}, image editing \cite{RN335, RN257, RN393}, and image reconstruction \cite{RN281, RN478, RN479}. In particular, DDPMs for text-to-image generation, such as Imagen \cite{RN214}, Stable \cite{RN227}, and DALL-E \cite{RN216}, garnered significant attention due to their ability to produce high-quality images with broad applicability. Although DDPMs excel in generating high-quality data, their main drawback is the need for a high computational cost to generate new samples, i.e., inference time, after the DDPMs are trained. DDPM is categorized as one of the probabilistic models that demands substantial computational resources. This complexity intensifies, particularly when handling high-dimensional data such as histopathological Whole Slide Images (WSI) \cite{RN468}. These images require significant memory resources and training time due to their exceptionally high resolution and large file size. Prominent advancements in diffusion models, notably demonstrated by Meta AI \cite{RN231} and Google Research \cite{RN204}, are attributed to their substantial computational capabilities. Evaluating pre-trained models entails significant time and memory costs, as the model may need to execute multiple steps to generate a data sample \cite{RN199}. This drawback poses a challenge for integrating these models into scenarios that demand quick and responsive results. For this reason, a new trend is to focus on improving the speed and scalability of diffusion models. 

Rombach et al. \cite{RN199} introduced the Latent Diffusion Model (LDM) to improve sampling efficiency and manage computational resources without sacrificing quality. It utilizes the latent space of a well-established pre-trained autoencoder, which offers a lower-dimensional representation that remains closely aligned with the data space. Consequently, it eliminates the need for spatial compression by training the diffusion model directly in the latent space. This approach exhibits superior scalability with respect to spatial dimensionality. Moreover, it strikes a balance between simplicity and the preservation of details, improving visual fidelity. Latent diffusion has been widely applied in generation tasks, demonstrating competitive performance. For instance, Pinaya et al. \cite{RN294} used LDM for synthesizing images from high-resolution 3D brain scans. Additionally, it is extensively used in image editing and reconstruction \cite{RN334}, classification \cite{RN446}, and image translation \cite{L1}. LDM substantially enhances both the training and sampling efficiency of DDPMs, while maintaining their high-quality outcome. Despite its success, there remains rooms to improve the inference speed. More recently, the Wavelet Diffusion Model (WDM) has been introduced to further improve the computational efficiency. WDM diffuses information across scales with wavelet transformations, allowing them to leverage parallel processing to reduce the computational cost required for training and generation of new synthetic data. WDMs also has demonstrated effectiveness in scalability, reflecting their ability to efficiently handle larger datasets without significant performance degradation. These attributes make diffusion models highly attractive from various perspectives \cite{RN197, RN202}. Incorporating wavelet transformation into the diffusion process improves image analysis \cite{RN233} and is widely used in various applications, such as image synthesis \cite{RN249, RN481}, image reconstruction \cite{RN482, RN253}, and image restoration \cite{RN478}. 
\begin{figure}[t]
  \centering

  \includegraphics[width=\linewidth]{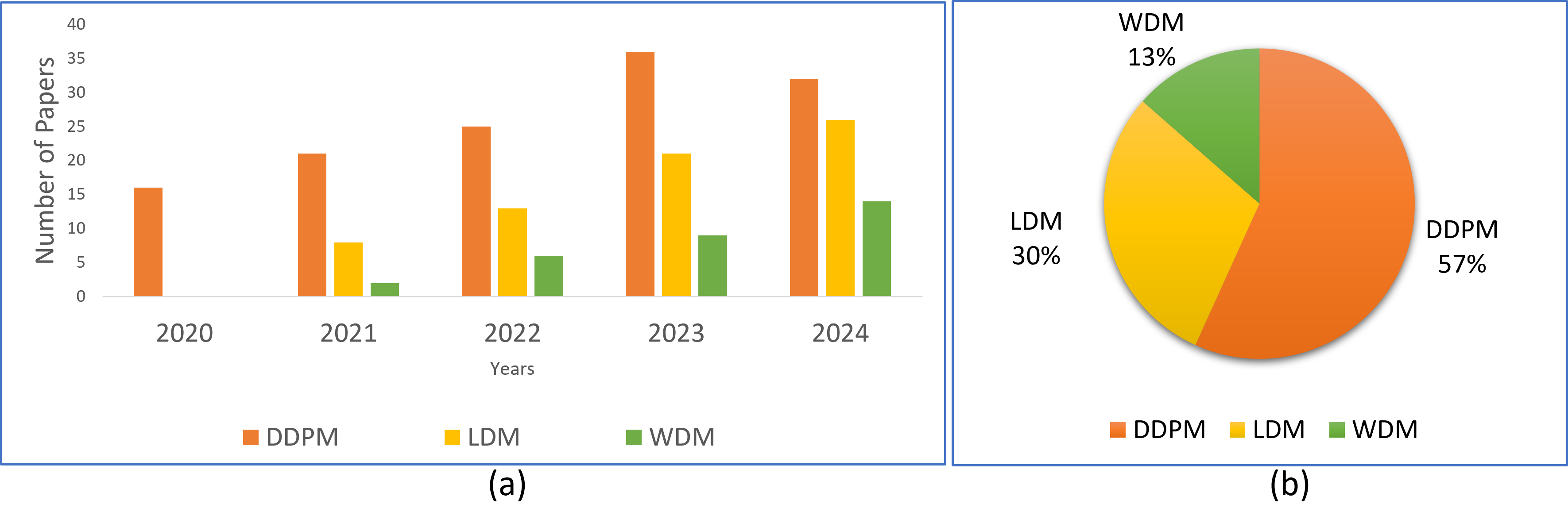}
  \label{fig:left}
\caption{ This reflects the evolving trends in landscape research. (a) The annual number of papers published related to DDPM, LDM, and WDM; (b) A total of 229 papers were identified. Within these categories, 130 papers focus on DDPM, 68 papers on LDM, and 31 papers on WDM.}
\label{fig:1}
\medskip
\Description{This reflects the evolving trends in landscape research.}
\end{figure}

\newcommand*\emptycirc[1][1ex]{\tikz\draw (0,0) circle (#1);} 
\newcommand*\halfcirc[1][1ex]{%
  \begin{tikzpicture}
  \draw[fill=blue] (0,0)-- (90:#1) arc (90:270:#1) -- cycle ;
  \draw (0,0) circle (#1);
  \end{tikzpicture}}
\newcommand*\fullcirc[1][1ex]{\tikz\fill[blue] (0,0) circle (#1);}
\setlength{\tabcolsep}{2pt}
\renewcommand{\arraystretch}{1.5}
\begin{table*}
\caption{Comparison of our survey paper with existing surveys on the diffusion model.}
\centering
\scriptsize 

\begin{tabular}{>{\raggedright\arraybackslash}p{0.8cm} >{\raggedright\arraybackslash}p{1cm} >{\raggedright\arraybackslash}p{2.4cm} >{\raggedright\arraybackslash}p{1cm} >{\raggedright\arraybackslash}p{1cm} >{\raggedright\arraybackslash}p{1cm} >{\raggedright\arraybackslash}p{0.8cm} >{\raggedright\arraybackslash}p{1cm} >{\raggedright\arraybackslash}p{1.3cm} >{\raggedright\arraybackslash}p{1cm} >{\raggedright\arraybackslash}p{1cm}} 
    \multirow{2}{*}{Surveys} & \multirow{2}{*}{Year} & \multirow{2}{*}{Survey Topics} & \multirow{2}{*}{Efficiency} & \multirow{2}{*}{Stability} & Fast Converge & Image Quality & Efficient Sampling & Generating Large Images & \multicolumn{2}{c}{Application} \\
    \cline{10-11}
    & & & & & & & & & Natural Image & Medical Image \\
    \midrule
    \cite{RN195} & 2022 & An Overview on Optimized Diffusion Model for Visual Task. & \centering\raisebox{-2ex}\fullcirc & \centering\raisebox{-2ex}\halfcirc & \centering\raisebox{-2ex}\halfcirc & \centering\raisebox{-2ex}\fullcirc & \centering\raisebox{-2ex}\emptycirc & \centering\raisebox{-2ex}\emptycirc & \centering\raisebox{-2ex}\halfcirc & \hspace{0.2cm}\raisebox{-2ex}\emptycirc \\
    \hline
    \cite{RN217} & 2023 & A Detailed Examination on Use of Diffusion Models in Medical Imaging. & \centering\raisebox{-2ex}\halfcirc & \centering\raisebox{-2ex}\halfcirc & \centering\raisebox{-2ex}\halfcirc & \centering\raisebox{-2ex}\halfcirc & \centering\raisebox{-2ex}\emptycirc & \centering\raisebox{-2ex}\halfcirc & \centering\raisebox{-2ex}\emptycirc & \hspace{0.2cm}\raisebox{-2ex}\fullcirc \\
    \hline
    \cite{RN192} & 2023 & A Review of Diffusion Model Applied to Vision System. & \centering\raisebox{-2ex}\halfcirc & \centering\raisebox{-2ex}\halfcirc & \centering\raisebox{-2ex}\emptycirc & \centering\raisebox{-2ex}\fullcirc & \centering\raisebox{-2ex}\emptycirc & \centering\raisebox{-2ex}\halfcirc & \centering\raisebox{-2ex}\fullcirc & \hspace{0.2cm}\raisebox{-2ex}\halfcirc \\
    \hline
    \cite{RN193} & 2023 & Through Overview of Diffusion Models Techniques and Use Cases. & \centering\raisebox{-2ex}\halfcirc & \centering\raisebox{-2ex}\emptycirc & \centering\raisebox{-2ex}\halfcirc & \centering\raisebox{-2ex}\halfcirc & \centering\raisebox{-2ex}\halfcirc & \centering\raisebox{-2ex}\emptycirc & \centering\raisebox{-2ex}\halfcirc & \hspace{0.2cm}\raisebox{-2ex}\halfcirc \\
    \hline
    \cite{huang2024diffusion} & 2024 & A Survey on Diffusion Model Approaches in Image Editing. & \centering\raisebox{-2ex}\halfcirc & \centering\raisebox{-2ex}\emptycirc & \centering\raisebox{-2ex}\halfcirc & \centering\raisebox{-2ex}\halfcirc & \centering\raisebox{-2ex}\emptycirc & \centering\raisebox{-2ex}\halfcirc & \centering\raisebox{-2ex}\halfcirc & \hspace{0.2cm}\raisebox{-2ex}\emptycirc \\
    \hline
    \cite{cao2024survey} & 2024 & Exploring Generative Diffusion Models: A survey. & \centering\raisebox{-2ex}\fullcirc & \centering\raisebox{-2ex}\halfcirc & \centering\raisebox{-2ex}\halfcirc & \centering\raisebox{-2ex}\emptycirc & \centering\raisebox{-2ex}\fullcirc & \centering\raisebox{-2ex}\emptycirc & \centering\raisebox{-2ex}\halfcirc & \hspace{0.2cm}\raisebox{-2ex}\halfcirc \\
    \hline
    \textbf{Our Survey} & 2024 & Computationally Efficient Diffusion Models in Medical Imaging: A Comprehensive Review. & \centering\raisebox{-2ex}\fullcirc & \centering\raisebox{-2ex}\fullcirc & \centering\raisebox{-2ex}\fullcirc & \centering\raisebox{-2ex}\fullcirc & \centering\raisebox{-2ex}\fullcirc & \centering\raisebox{-2ex}\fullcirc & \centering\raisebox{-2ex}\fullcirc & \hspace{0.2cm}\raisebox{-2ex}\fullcirc \\
    \bottomrule
  \end{tabular}
\\[1ex]
\textbf{Note:}\hspace{0.4em}{The \hspace{0.4em}\raisebox{-0.5ex}\fullcirc \hspace{0.4em} indicates that all aspects are covered, the \hspace{0.4em}\raisebox{-0.5ex}\halfcirc\hspace{0.4em} shows that certain aspects are partially addressed, and the \hspace{0.4em}\raisebox{-0.5ex}\emptycirc\hspace{0.4em} signifies that no discussion has occurred.}
\label{fig:table1}
\end{table*}
In this paper, we present a comprehensive overview of the recent advances of  DDPM, LDM, and WDM, highlighting their respective applications in both computer vision and medical image analysis. Fig. \ref{fig:1} illustrates the emerging research trends in diffusion generative models, focusing on three main models based on their generative capabilities and the efficiency of inference and training times. Specifically, Fig. \ref{fig:1}(a) presents the annual publication count for these models, while Fig. \ref{fig:1}(b) highlights the most widely adopted model in the current research field. 
While there are existing surveys on diffusion models and their applications, no existing studies specifically discuss the latency and time complexity of generative models, specifically dealing with extremely large high-resolution data such as WSI or 3D MRI/CT. This survey highlights this gap for further research advancements. Table \ref{fig:table1} provides a comparative analysis between our work and other surveys. We explore various aspects of DDPM, LDM, and WDM, including performance, efficiency, and stability, relative to existing diffusion models. Furthermore, we discuss potential applications, benefits, and challenges of using these models for medical image analysis and natural images. This study aims to provide valuable insights to the research community, paving the way for advancements in harnessing the potential of these models in medical imaging. Croitoru et al. \cite{RN192} provided a survey article to the utilization of diffusion models in computer vision. The article highlights various diffusion models and compares them with existing generative models based on both generative quality and computational cost. Similarly, Yang et al. \cite{RN193} discussed an extensive detail of current advancement in diffusion models. This study identifies a comprehensive perspective on improving diffusion algorithms and exploring their applications across multiple domains, such as text generation, image synthesis, video generation, and music composition. More relevant to our review is the research conducted by Kazerouni et al. \cite{RN217}, which underscores the role of the diffusion model in medical image analysis. This article explicitly focuses on the utilization of the diffusion model in medical applications, considering key factors such as modality, organ of interest, and algorithms used. However, they did not specifically discuss latency and execution time, which currently stands as one of the most challenging aspects of generative artificial intelligence. 

We have highlighted the key findings and outcomes of this review paper outlined as follows:
\begin{itemize}[left=1em]
   \item We examine the fundamental concepts of DDPM, LDM, and WDM, shedding light on their respective strengths and limitations.
   \item We explain the significance of DDPM, LDM, and WDM across various applications, particularly in natural and medical imaging analysis.
   \item We underscore the pivotal role of the WDM and LDM in addressing the trilemma problem inherent in diffusion models, offering valuable insights such as the generation of high-quality samples, accelerated sampling process, and improved mode convergence.
   \item Our survey paper distinguishes itself by highlighting the unique characteristics within the landscape of existing surveys on diffusion models shown in Table \ref{fig:table1}. In addition, we provide a critical analysis corresponding to each section for a comprehensive evaluation and insight.
  \item Additionally, we identify and articulate the limitations and open gaps in current research, offering directions for future exploration and advancement within the research community. 
\end{itemize}
The remaining sections of this survey article are structured as follows: Section 2 provides an overview of diffusion models, explicitly emphasizing the three models DDPM, LDM, and WDM. Section 3 delves into multipurpose applications, focusing on the medical domain. Section 4 conducts a comparative analysis, evaluating both quality and computational cost. Section 5 offers an overview of future prospects and challenges in the field. Section 6 provides a concluding discussion.
\section{Preliminary: Diffusion Models Concept}\label{sec:diffusion}
Diffusion models were inspired by the principles of non-equilibrium irreversible thermodynamics. This theoretical approach encapsulates the irreversibility inherent in a dynamic system, where changes are observable in a manner that deviates from perfect reversibility.
\subsection{Diffusion Models}
In this study, we focus on three models, including DDPM \cite{RN224}, LDM \cite{RN199} and WDM \cite{RN233} based on high-quality images generative capabilities while addressing critical challenges of trilemma problem \cite{RN260}.
\subsubsection{Foundation of Denoising Diffusion Probabilistic Model (DDPM)}
\begin{figure}[t]
\medskip
    \centering
    \includegraphics[width=\linewidth]{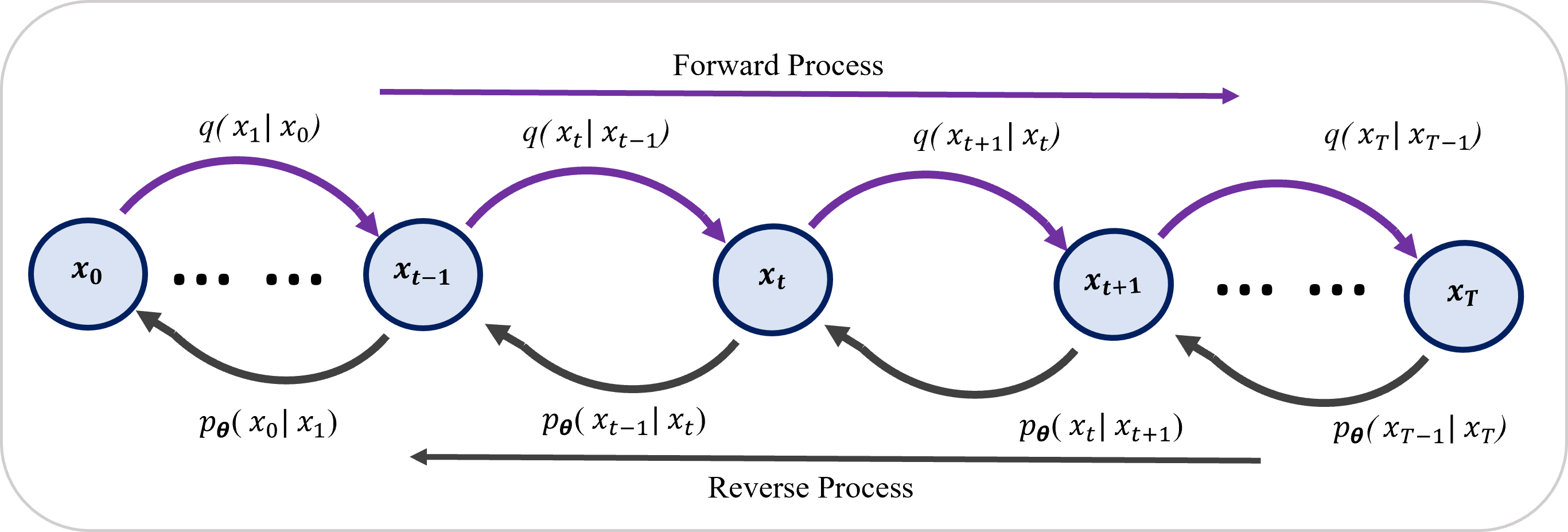}
    \caption{An Illustration of the DDPM forward process \(q(x_t| x_{t_1})\) add noise and reverse process \(p_{\theta}(x_{T-1}|x_t)\) removing noise.}
    \label{fig:2_1}
    \medskip
    \Description{An Illustration of the DDPM forward process and reverse process}
\end{figure}

DDPM \cite{RN224} employs two Markov chains, encompassing both forward and reverse diffusion processes. The forward diffusion process involves gradually injecting random noise to perturb complex image data in many steps. This iterative exploration involves incremental adjustments to adapt to the data distribution. The noise scale varies at each step, contributing to the progressive transformation of the data. The same iterative procedure is employed to retrieve the original image by removing the noise from the noisy image \cite{RN199, RN224}; consider a forward diffusion process involving $k$ timesteps designed to gradually perturb each input original data sample denoted as \(x_0\), following an accurate data distribution \(q(x_0)\) as shown in Fig. \ref{fig:2_1}, where $q$ shows a probability function. In this process, Gaussian noise is incrementally added in sequential steps, resulting in distorted samples. The generation of latent variables \(x_1, x_2, \dots, x_{S-1}, x_S\) involves the addition of Gaussian noise at each timestep $t$.
The conditional probability distribution \( q(x_t \mid x_{t-1}) \) is given by:
\begin{align*}
    q(x_t \mid x_{t-1}) &= \mathcal{K}\left(x_t; \sqrt{1-\beta_t}x_{t-1}, \beta_t \mathit{I}\right)\textcolor{red}{,} 
    &\text{such that } t \in \{1, \dots, S\}.\ 
    \label{eq:1}
    \tag{1}
\end{align*}

Meanwhile, $S$ represents the total number of sequences in the diffusion process. The hyperparameter \(\beta\) characterizes the variance across various diffusion steps. The identity matrix, denoted as \(I\), has the same dimensions as the original data \(x_0\). The distribution \(\mathcal{K}(x, \mu, \sigma)\) reflects a normal distribution with mean \(\mu\) and variance \(\sigma\), where  \(\mu\)\ signifies the mean and \(\sigma\)\ represents the variance of the distribution. This normal distribution is applied to perturb the data at each diffusion step. A variance schedule manages the step size across the diffusion process stages. This schedule determines how the variance of the added Gaussian noise evolves throughout diffusion steps, influencing the extent of perturbation at each timestep \( \beta_t \) satisfies \( \beta_t \in (0,1) \) for \(t \in \{1, \dots, S\}\). Considering \( \alpha_t = 1 - \beta_t \), and \( \overline{\alpha}_t = \prod_{s=0}^{t} \alpha_s \), whereas  \( \alpha_t\)\ represents the complementary probability of  \( \beta_t \), while \(\overline{\alpha}_t\) represents the cumulative product of \( \alpha_s\)\ from  \(s\) = 0 to \(t\).
The posterior probability of a diffusion image \(x_t\) at timestep \(t\) can be formulated in a closeform expression as follows:  
\begin{equation}
    q(x_t \mid x_0) = \mathcal{K}(x_t; \sqrt{\overline{\alpha}_t x_0}, (1 - \overline{\alpha}_t)I).\
    \tag{2}
    \label{eq:2}
\end{equation}
Here, \(x_{0}\) represents the original input images, and the diffusion process to obtain \(x_t\)  involves sampling a Gaussian vector \( \epsilon \sim \mathcal{K}(0, 1) \) and applying a transformation. This forward process gradually introduces noise to the data, leading to the loss of structured information over time. In generating new data samples, DDPMs imitate the process by generating a prior-distribution noise vector with no inherent structure, which is typically easy to obtain. 
\begin{equation}
    x_t = \sqrt{\overline{\alpha}_t x_0} + \sqrt{1 - \overline{\alpha}_t} \epsilon.\ \tag{3}
    \label{eq:3}
\end{equation}
The subsequent step involves progressively removing noise by employing a trainable Markov chain designed to run in reverse temporal order. This reverse diffusion process aims to reconstruct structured information from the initially introduced noise, ultimately generating realistic data samples. The iterative removal of noise allows the model to capture and recreate meaningful structures present in the original data, generating novel and coherent samples. The reverse conditional probability, often denoted as \(p_{\theta}(x_{t-1} \mid x_t) \), can be traced as shown in Fig. \ref{fig:2_1},  when condition on \( x_{t} \) a perturbated image that maintains an identical function structure to the forward process \(q(x_t \mid x_{t-1})\)[39].
\begin{equation}
 p_{\theta}(x_{t-1} \mid x_t) = \mathcal{K}(x_{t-1}; \mu_0(x_{t}, t), \beta^2 I).\
 \tag{4}
 \label{eq:4}
\end{equation}
In an ideal neural network training, maximizing likelihood means maximizing the probability \(p_{\theta}(x_0)\), assigned to each training sample \(x_0\) by the denoising module \(\mu_0(x_{t},t)\). However, calculating \(p_{\theta}(x_{0})\) is intractable due to the need to marginalize over all possible reverse trajectories. The primary objective is to reduce the Kullback-Leibler \((KL)\) divergence between the actual denoised data distribution and the distribution represented by the model.
\begin{equation}
L_{\text{vlb}} = -\log p_{\theta}(x_0 \mid x_1) + \textit{KL}(p_{\theta}(x_S \mid x_0) \Vert \pi(x_S) + (x, z)) + \log p_{\theta}(x)\textcolor{red}{.} \tag{5}
\label{eq:5}
\end{equation}
\subsubsection{Latent Diffusion Model (LDM)}
\begin{figure}[t]
\medskip
    \centering
\includegraphics[width=\linewidth]{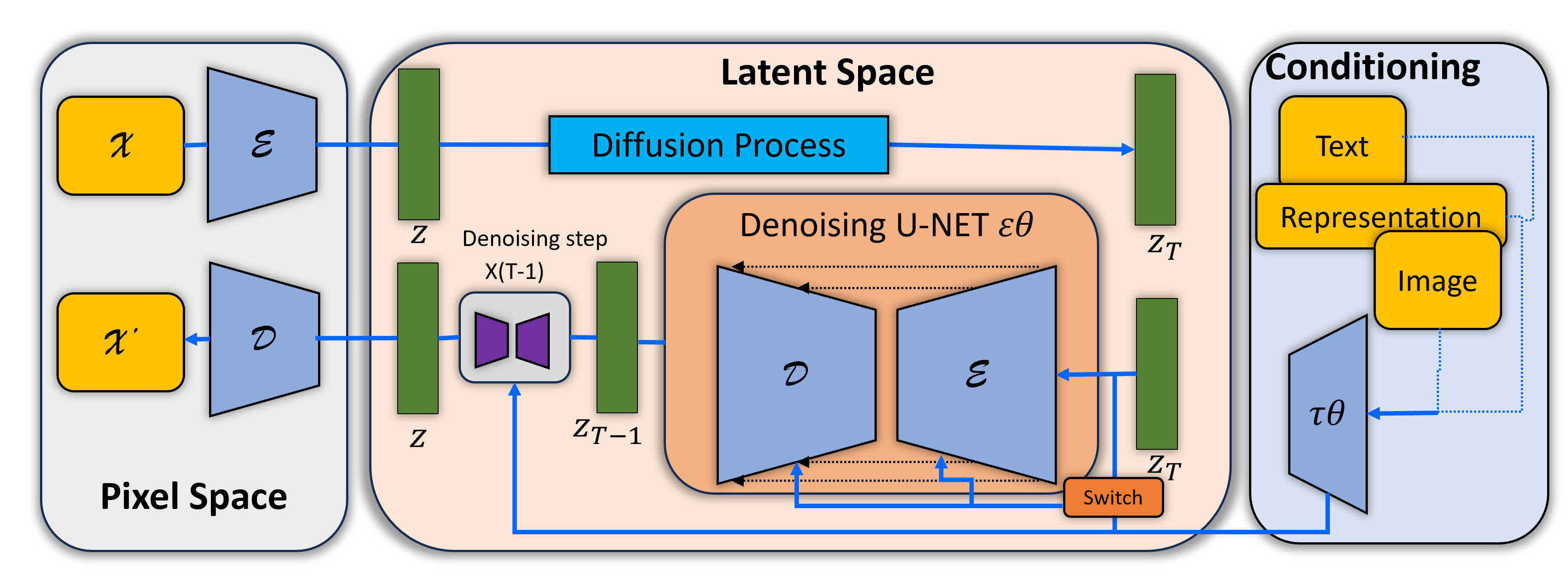}
    \caption{LDM basic framework encodes data from the pixel space into a latent space. Within this latent space, the diffusion process is carried out, guided by the conditional input modality for synthesis. Finally, the model decodes the data back into the image domain \cite{RN199}.}
    \label{fig:LDM_2}
    \medskip
    \Description{LDM}
\end{figure}

The process of training diffusion models directly in pixel space for high-resolution image synthesis poses significant computational challenges. This complexity arises from the high dimensionality inherent in pixel-based representations, which require substantial processing power, and memory resources. To address this challenge, LDM \cite{RN199} utilizes a pre-trained encoder to project the image into a lower-dimensional space. This projection allows the diffusion process to occur within a compressed latent space, which considerably reduces the computational demands as shown in Fig. \ref{fig:LDM_2}. As the diffusion process operates on a more compact representation, it requires fewer operations at each step, facilitating faster training and inference time. The LDM is consistent with that of traditional diffusion models which are designed 
 to learn from a data distribution  \(q(x)\) in a pixel space. This is achieved by training a denoising autoencoder to predict a denoised variant of the input \(x_t\), where \(x_t\) denotes a noisy version of the input \(x\).

\begin{equation}
\mathcal{L}_{\text{DM}} = \mathbb{E}_{\mathbf{\epsilon{(x)}},\varepsilon \sim \mathcal{N}(0, 1),t} \left[ \left\| \varepsilon - \varepsilon_{\theta}(\mathbf{x}_t, t) \right\|^2_2 \right] {.} \tag{6}
\label{eq:6}
\end{equation}

\begin{equation}
\mathcal{L}_{\text{LDM}} = \mathbb{E}_{\mathbf{\epsilon{(x)}}, \varepsilon \sim \mathcal{N}(0, 1),t} \left[ \left\| \varepsilon - \varepsilon_{\theta}(\mathbf{z}_t, t) \right\|^2_2 \right] {.} \tag{7}
\label{eq:7}
\end{equation}
However, this compression model consists of an encoder \( \epsilon\) and neural backbone \(\varepsilon \), which work together to produce a compact latent space \(z_t\). It effectively eliminates the high-frequency details during the training of \(\epsilon\)  from the data distribution \(q(x)\). This low-dimensional representation facilitates a concentrated focus on essential semantic information while enhancing model training efficiency through dimensionality reduction. Furthermore, various generative models \cite{L3} possess the capability to model the conditional distribution \(q(z/x')\). This is especially pertinent in LDM, whereas conditional denoising autoencoder regulates the synthesis output based on a specified input condition \(x'\). These input conditions can be derived from various modalities, such as images and text. Accordingly, LDM employs a domain-specific encoder  \(\tau_{\theta}(x')\), which is mapped to the intermediate layers of U-Net. Thus, for image-conditioning pairs, LDM can be defined as follows: 
\begin{equation}
\mathcal{L}_{\text{LDM}} = \mathbb{E}_{\mathbf{\epsilon{(x)}},  x', \varepsilon \sim \mathcal{N}(0, 1)} \left[ \left\| \varepsilon - \varepsilon_{\theta}(\mathbf{z}_t, t,\tau_{\theta}(x) ) \right\|^2_2 \right] {.} \tag{8}
\label{eq:8}
\end{equation}

Equation \eqref{eq:8} indicates that the neural network \(\varepsilon_{\theta}\) learns and optimizes by incorporating both the latent space parameters \(z_t\) and the conditions inputs \(\tau_{\theta}(x')\). To enhance the understanding of this process, Lin et al. \cite{L1} developed a Syncretic LDM \((S^2LDM)\), which uses non-contrast CT (NCCT) as the source and contrast-enhanced CT (CECT) image as the target. They utilized a dynamic similarity mask as a condition to guide the model convergence specifically in the contrast-enhanced regions. Additionally, Takagi et al. \cite{L4} introduced a method to reconstruct high-resolution brain images using functional MRI (fMRI), whereas fMRI is a neuroimaging technique that captures anatomical images of brain structure while reflecting the dynamic changes in brain activity. The text associated with the fMRI data is used as a condition to reconstruct high-fidelity images with reduced complexity. 
\subsubsection{Wavelet Diffusion Model (WDM)}
Diffusion models are renowned for their ability to produce high-quality, stable outputs, but their lengthy inference times present challenges for real-time applications. 
Recent advancements have leveraged the wavelet transform \cite{RN239}, a widely adopted image compression and reconstruction technique, to enhance computational efficiency and feature interpretability within diffusion models. The wavelet transform's ability to decompose images into multiscale components has proven invaluable, enabling more effective representation and manipulation of image data. 
\begin{figure}[t]
    \centering
    \includegraphics[width=1\textwidth]{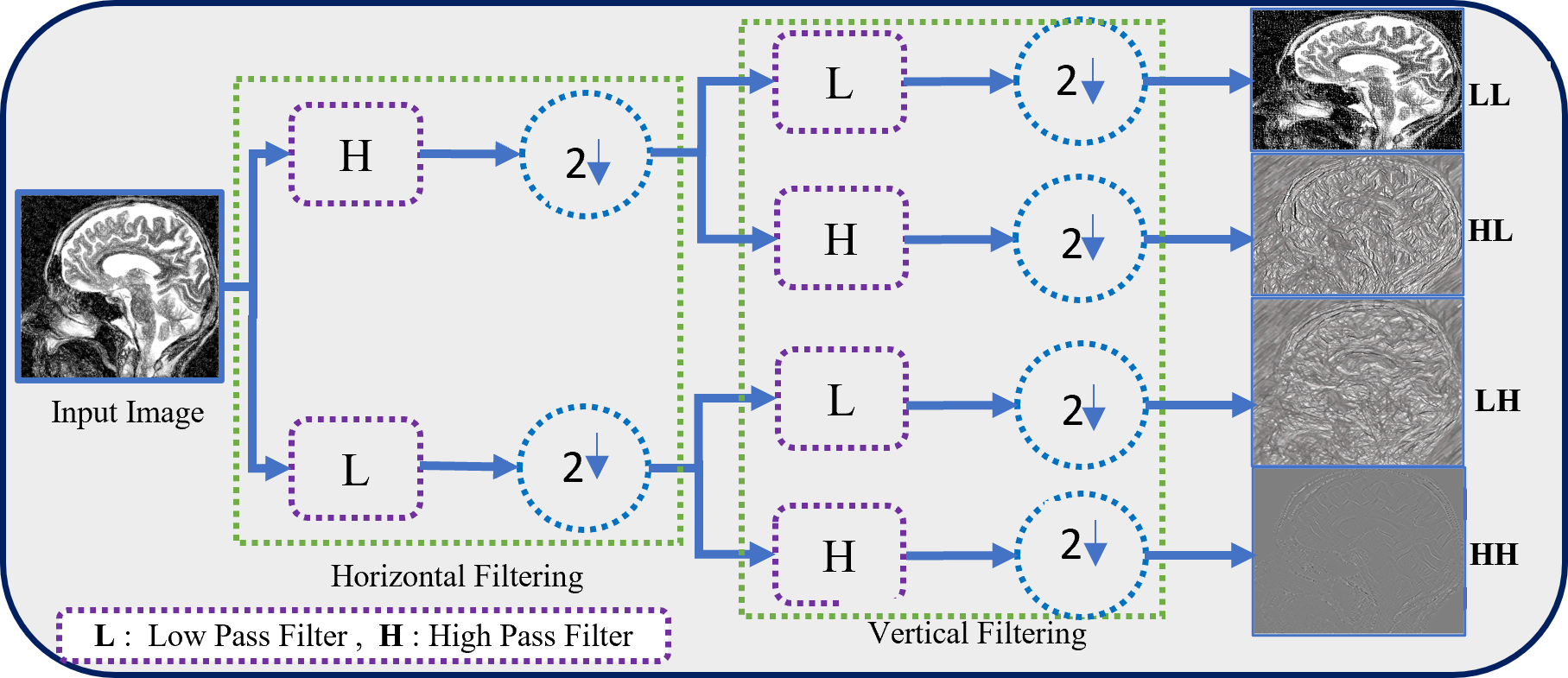}
    \caption{The wavelet transform method decomposes an image into its frequency components at different scales to capture both contextual and wide-range features efficiently: Implementation of 2D-DWT.}
    \label{fig:3}
    \Description{Wavelet Transformation}
\end{figure}
The Discrete Wavelet Decomposition (DWT) and its Inverse Wavelet Decomposition (IWT) can perform by using low and high pass filters as follows: ${L} = \frac{1}{\sqrt{2}} \begin{bmatrix} 1 \\ 1 \end{bmatrix}$ and  ${H} = \frac{1}{\sqrt{2}} \begin{bmatrix} -1 \\ 1 \end{bmatrix}$. These two filters involve the construction of four kernels, namely ${LL, LH, HL, HH}$. This operation facilitates the decomposition of the input image into four subbands, resulting in subbands with dimensions $L/2$ and $H/2$. The transformed low-frequency subband ${LL}$ preserves essential image characteristics, including energy, while the high-frequency subbands ${LH, HL, HH}$ exhibit strong spatial selectivity, capturing detailed information shown in Fig. \ref{fig:3}. The reconstruction process uses the same structure to reconstruct the original image. This IWT  method efficiently maintains global and local image features, making it valuable for image analysis and compression \cite{RN233, RN386}. Integrating wavelet transform into a diffusion process involves concatenating decomposed subbands into a unified target for denoising. The model operates in the wavelet spectrum rather than spatial image space, leveraging high-frequency information from subbands to enhance image details. Additionally, the smaller spatial area of wavelet subbands (one-fourth of the original images) significantly reduces computational complexity during sampling. However, the DD-GAN model \cite{RN260} which combines the rapid sampling capabilities of GANs with the superior sampling-quality outputs of diffusion models, serves as a foundational model in this approach. Mathematically integrating GAN  generator \(G_\phi\) and a discriminator \(D_\phi\) aiming to model the data distribution \(p{\theta}(x)\). Here, the generator produces realistic samples \(G_\phi(z)\) from random noise \(z\), while the discriminator distinguishes between real data \(x\) and generated samples \(G_\phi(z)\). GANs use a min-max objective function where \(G_\phi\) tries to deceive \(D_\phi\), and \(D_\phi\) aims to classify real and fake samples accurately. This adversarial process leads to \(G_\phi\) capturing the underlying data distribution. Like traditional diffusion models, Diffusion GAN facilitates sampling speed with large step size, using adversarial training for the discriminator as defined in the equation below:
\begin{equation}
\min_{\Phi}\max_{\theta} \left( \sum_{t\geq1} \mathbb{E}_{q(x_t)} \left[ \mathbb{E}_{q(x_{t-1} | x_t)} \left[-\log \left( D_{\phi}(x_{t-1}, x_t; t) \right) \right] + \mathbb{E}_{p{\theta}(x_{t-1} | x_t)} \left[-\log \left( D_{\phi}(x_{t-1}, x_t; t) \right) \right] \right] \right).  \tag{9}
\label{eq:9}
\end{equation}
The generator is trained by contrasting fake samples from the distribution \( p_{\theta}(x_{t-1}, x_t) \) with real samples from \(q(x_{t-1}, x_t)\). The challenges lie in the fact that the first expectation involves sampling from the unknown distribution \(q(x_{t-1}, x_t)\). In equation \ref{eq:9}, the original input image is represented as \(x_0\), and \(x_t\) denotes noisy samples obtained by adding Gaussian noise at successive timestep \(t\), formulated as \(q(x_t\mid x_0)\). The latent variable $z$ is generated to follow a normal distribution $N(0,1)$. At timestep $t$, the generated original image sample is represented as $x_0^{\prime}=G(x_0, z, t)$. Subsequently, the estimated noisy sample $x_{t-1 }^{\prime}$ is obtained from the posterior distribution $q(x_{t-1}\mid x_t, x_0^{\prime})$. The discriminator distinguishes the real $(x_{t-1}\mid x_t)$ and fake  $(x_{t-1 }^{\prime}\mid x_t )$ pairs \cite{RN233}. 
Recently, studies explored wavelet generative models, including the Wavelet Score-Based Model by Guth et al. \cite{RN256}. This model synthesizes coefficients with a consistent number of time steps at all scales. While the resource demands scale linearly with image size, normalization enables a uniform discretization schedule. Consequently, the total sampling iterations per image remain independent of its size. Jiang et al. \cite{RN257} proposed a wavelet-based conditional diffusion model for enhancing low-light images, integrating a high-frequency restoration module to capture diagonal details for improved image fidelity. The model employs DWT to decompose images into frequency bands, where it learns the low-frequency spectrum of clean images by introducing stochastic noise during training. The integration of the low-frequency band with the high-frequency enhancement module is achieved through IWT \cite{RN258}, allowing for efficient recovery of degraded images by leveraging both spatial and frequency domains
\subsection {Critical Analysis and Discussion}
Major challenge in current models is balancing image quality with processing speed. Higher image quality often slows down training and inference process, while faster processing can compromise quality. In medical imaging, quality is essential without incurring high processing costs, whereas in natural imaging, particularly video generation, inference speed is prioritized, though quality remains important. We analyze and draw conclusions based on the characteristics highlighted in Table \ref{fig:table1} concerning DDPM, LDM, and WDM.
\subsubsection{Efficiency:} DDPM removes noise through a slow, iterative process, ensuring high accuracy but resulting in longer training times compared to LDM. In contrast, LDM operates within a latent space, significantly enhancing training speed and overall efficiency. WDM introduces a more efficient alternative method by decomposing images into frequency components, thus accelerating both training and inference while capturing finer details. Ultimately, the efficiency of these models varies based on specific conditions and task requirements. 
\subsubsection{Stability:} DDPM, is recognized for its stability in image generation due to a systematic noise reduction process. However, their generation speed is often compromised due to the extensive diffusion steps required, which can hinder the preservation of details particularly in medical images. LDM executes the diffusion process in a lower-dimensional latent space, and it provides better stability based on latent space conditions because its generated output depends on latent space. WDMs enhance stability by focusing on multi-resolution details, promoting controlled sampling. However, WDM's robustness can be limited compared to LDM, making its stability dependent on the employed regularization techniques and specific tasks.
\subsubsection{Fast Convergence:} DDPM necessitates numerous sampling steps to produce high-quality images, resulting in slower convergence. In contrast, LDM achieves faster convergence by utilizing a lower-dimensional latent space, having a balance between convergence speed and the quality of generated images. WDM further enhances convergence speed by integrating wavelet-based regularization, enabling quicker capture of both global and local image features.
\subsubsection{Image Generation Quality:} When evaluating generation quality, DDPM excels at producing high-fidelity, diverse images, resulting in intricate and realistic outputs. LDM also delivers high quality using latent space but loses details, especially when generating high-resolution images. In contrast, WDMs demonstrate enhanced performance in texture diversity, yielding robust outputs. Collectively, these models present a comparative landscape: DDPM stands out for overall image quality, LDMs provide high quality while less than DDPM, and WDMs excel in texture diversity and details preservation, making them suitable for specific applications, such as medical image generation.
\subsubsection{Efficient Sampling:} 
In terms of efficient sampling, DDPM is characterized by low efficiency due to slow sampling resulting from numerous diffusion steps. Conversely, LDM achieves high efficiency through sampling in a lower-dimensional latent space, greatly speeding up the process. While, WDM offers moderate efficiency, as their multi-resolution sampling approach may hinder overall speed. LDMs are recommended for applications requiring a balance of quality and efficient sampling. Meanwhile, WDMs can be effective in sampling, possess more complex implementation
\subsubsection{Generating Large Images:}
Scalability is a notable challenge for DDPM, which can become computationally inefficient when generating larger images. In contrast, LDMs perform better with large images, although performance heavily depends on the choice of latent space during model pre-training. However, WDM provides a more effective solution by decomposing images into distinct frequency bands, thereby improving scalability and preserving high-frequency details essential for large images. Generally, LDMs are considered more efficient for large image generation due to their lower processing requirements and ability to maintain image quality. However, if preserving fine details is critical, WDMs may be a suitable choice.
\subsubsection{Application in Natural and Medical Imaging:} DDPMs are effective for generating natural images due to their flexibility and robustness. However, they are less suited for medical imaging due to their limited interpretability, a crucial aspect of medical applications. LDMs perform well across diverse datasets, which solves the trilemma problem by balancing efficiency and generative quality, particularly in natural imaging. In contrast, Wavelet Diffusion excels in medical imaging by capturing large structures and intricate details, making it a valuable tool for diagnosis while remaining competitive in natural image generation.

Table \ref{table:2} compares the key characteristics of three models: DDPM, LDM, and WDM. Table \ref{table:3} provides a detailed comparative analysis of these models across three popular public benchmark datasets. It evaluates their generative quality using FID and Recall scores, while also examining computational complexity in terms of model parameters and inference time DDPM achieves slightly better FID scores on CIFAR-10 compared to LDM and WDM. WDM and LDM, however, offer faster inference times across datasets, highlighting a quality-speed trade-off. Ultimately, LDM balances quality and efficiency for general use, though WDM is a strong option if faster inference with minimal quality loss is required. Generally, models that ensure high-quality images with acceptable efficiency are valuable.
\begin{table}[t]
\centering
\setlength{\tabcolsep}{2pt}
\scriptsize 
\renewcommand{\arraystretch}{1.2} 
\setlength{\tabcolsep}{6pt} 
\caption{Comparison of key characteristics of WDM, LDM, and DDPM.}
    \begin{tabular}{p{1.25cm} @{\hspace{0.5cm}} p{1.5cm} @{\hspace{0.5cm}} p{1.5cm} @{\hspace{0.5cm}} p{1.5cm} @{\hspace{0.25cm}} p{2cm} @{\hspace{0.25cm}} p{1.2cm}@{\hspace{0.5cm}} p{1.8cm}}
    \hline
    \textbf{Models} & \textbf{Efficiency} & \textbf{Stability} & \textbf{Fast convergence} & \textbf{Image generation quality}& \textbf{Efficient sampling} & \textbf{Generating large image}\\
    \hline
    \text{WDM \cite{RN233}} & \text{High} & \text{Moderate} & \text{High} & \text{Moderate} & \text{Low} & \text{High} \\ 
    \text{LDM \cite{RN199}} & \text{Moderate} & \text{Moderate} & \text{High} & \text{Moderate} & \text{Moderate} & \text{Moderate} \\
    \text{DDPM \cite{RN224}} &\text{Low} & \text{High} & \text{Low} & \text{High} & \text{Low} & \text{Low}\\
    \hline
    \end{tabular}\\
    \medskip
    \label{table:2}
\end{table}
\begin{table}[htpb]
\centering
\medskip
\scriptsize 
\renewcommand{\arraystretch}{1.2} 
\setlength{\tabcolsep}{6pt} 
\caption{Performance evaluation of WDM, LDM, and DDPM for image synthesis.}
    \begin{tabular}{@{\hspace{1cm}} p{1.5cm} @{\hspace{1cm}} p{2.2cm} @{\hspace{0.6cm}} p{1.2cm} @{\hspace{0.6cm}} p{1.2cm} @{\hspace{0.6cm}} p{1.2cm} @{\hspace{0.6cm}} p{1.5cm}}
    \hline
    \textbf{Method} &\textbf{Dataset} & \textbf{FID $\downarrow$ } & \textbf{Recall $\uparrow$} & \textbf{Params $\downarrow$} & \textbf{Time(s) $\downarrow$}\\
    \hline
    \multirow{3}{*}{WDM \cite{RN233}} & CIFAR10 \cite{D27} & 4.01 & 0.55 & 33.37M & 0.08 \\ 
        & Celeba-HQ \cite{D17} & 6.55 & 0.35 & 44.85M & 0.60 \\
        & LSUN-Church \cite{D74} & 5.06 & 0.40 & 31.48M & 1.54\\
        \hline
        \multirow{3}{*}{LDM \cite{RN199}} & FFHQ \cite{D11} & 4.98 & 0.50 & 274M & 0.43 \\ 
        & Celeba-HQ \cite{D17}  & 5.11 & 0.49 & 274M & 1.07\\ 
        & LSUN-Chruch \cite{D74} & 4.02 & 0.52 & 256M & 6.80\\
         \hline
        \multirow{3}{*}{DDPM \cite{RN224}} & CIFAR10 \cite{D27}& 3.21 & 0.57 & - & 80.5 \\ 
         & Celeba-HQ \cite{D17} & 3.77 & - & - & - \\ 
        & LSUN-Church \cite{D74} & 7.89 & - & - & - \\
        \hline
    \end{tabular}\\
    \textbf{Note:}\hspace{0.2em} \text{FID:} Frechet Inception Distance, near to zero indicates high quality. \hspace{0.2em} \text{Time:} More efficient with a lower value.
    \medskip
    \label{table:3}
\end{table}

\section{Applications}
This section provides a broad overview of the applications of diffusion models in natural and medical image analysis.
\begin{figure}[htbp]
    \centering
    \includegraphics[width=\linewidth, height=0.8\textheight]{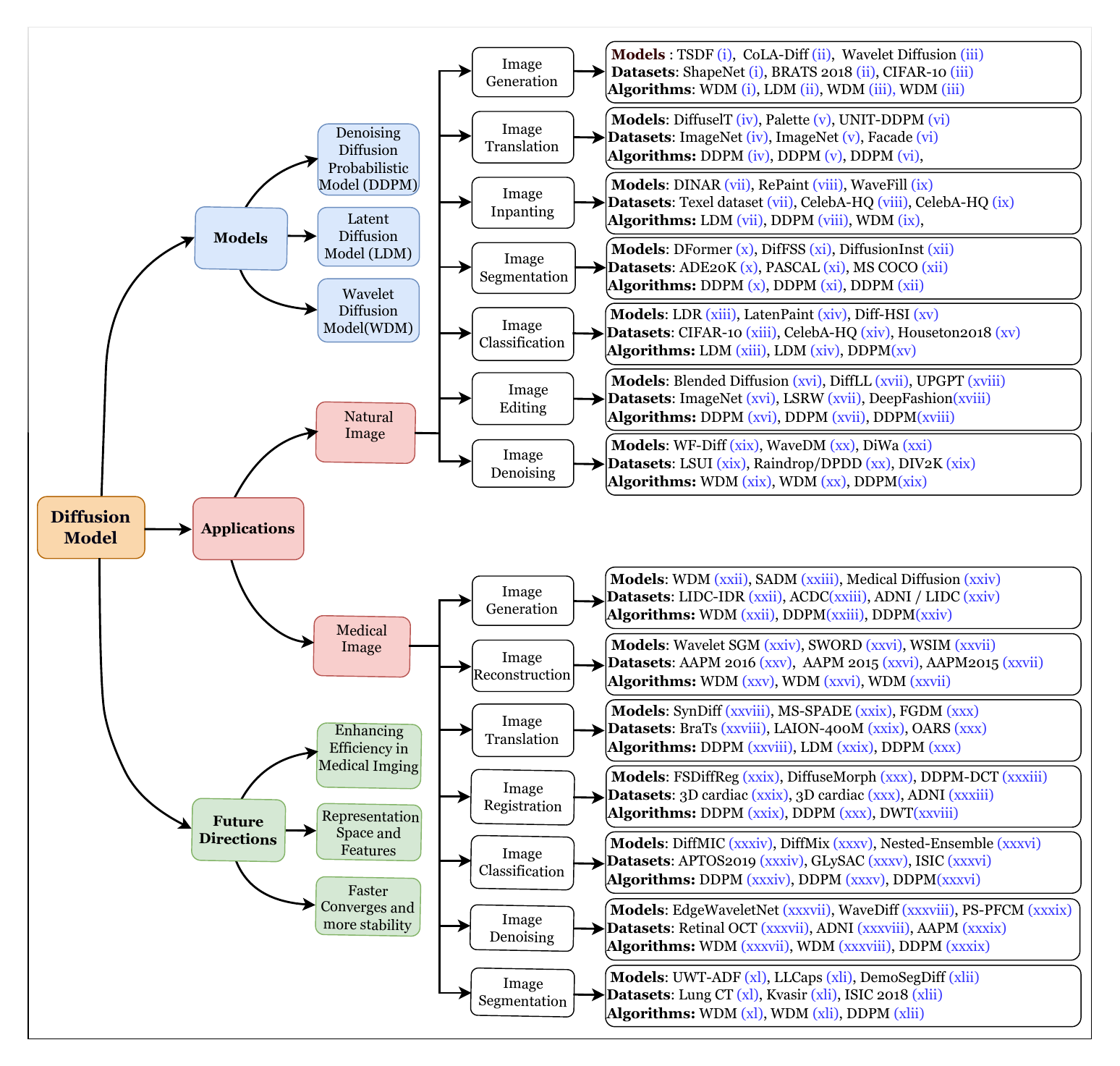}
    \caption{We categorize diffusion models into models, applications, and future directions. Notably, DDPM, LDM, and WDM are highlighted for their efficiency and stability. Applications are further categorized into two mainstream natural and medical Images with their specific tasks. For clarity, paper references are indicated by ascending prefix numbers: (i) \cite{RN259}, (ii) \cite{RN447}, (iii) \cite{RN233}, (iv) \cite{RN448}, (v) \cite{RN267}, (vi) \cite{RN327}, (vii) \cite{RN265}, (viii) \cite{RN266}, (ix) \cite{RN186}, (x) \cite{RN416}, (xi) \cite{RN417}, (xii) \cite{RN418}, (xiii) \cite{RN477}, (xiv) \cite{RN264}, (xv) \cite{RN427}, (xvi) \cite{RN335}, (xvii) \cite{RN257}, (xviii) \cite{RN393}, (xix)\cite{RN478}, (xx) \cite{RN258}, (xxi) \cite{RN479}, (xxii) \cite{RN480}, (xxiii) \cite{RN396}, (xxiv) \cite{RN445}, (xxv) \cite{RN481}, (xxvi) \cite{RN482}, (xxvii) \cite{RN483}, (xxviii) \cite{RN273}, (xxix) \cite{L5}, (xxx) \cite{RN389}, (xxxi) \cite{RN459}, (xxxii) \cite{RN269}, (xxxiii) \cite{RN484}, (xxxiv) \cite{RN270}, (xxxv) \cite{RN412}, (xxxvi) \cite{RN413}, (xxxvii )\cite{RN253}, (xxxviii) \cite{RN306}, (xxxix) \cite{RN464}, (xl) \cite{RN486}, (xli) \cite{RN487}, (xlii) \cite{RN401}.}
    \label{fig:taxnomy}
\Description{We categorize diffusion models into models, applications, and future directions. Notably, DDPM, LDM, and WDM are highlighted for their efficiency and stability. Applications are further categorized into two mainstream natural and medical Images with their specific tasks}
\end{figure}
\subsection {DDPM for Natural and Medical Image}
We divided the applications of DDPM into two subcategories natural and medical image analysis.
\subsubsection{\textbf{DDPM for Natural Image Analysis.}}
Figure \ref{fig:taxnomy} presents the overall structure of this survey paper, categorized into three main groups: models, applications, and future directions. The applications are further divided into natural and medical image categories. This section discusses the applications of DDPM in natural images.   
\subsubsubsection{Generation.}
Image generation is a primary goal for DDPM, extensively applied across various domains to synthesize the images \cite{RN228, RN393, RN396, RN391, RN392, RN395, RN397, RN243}. Epstein et al. \cite{RN400} introduced a self-guidance zero-shot approach that facilitates image generation without the need for prior training. This method enables precise control over the shape, position, and appearance of objects in generated images, offering flexibility in creative applications. Bar-Tal et al. \cite{RN397} introduced a multi-diffusion approach, which combines multiple diffusion processes under shared parameters to enhance attribute-specific image generation. This advancement enables text-to-image models for practical use, allowing users to adjust aspect ratios, control scene layouts, and generate text-to-panorama images.
\subsubsubsection{Translation.}
Transforming images between domains while preserving key characteristics is vital, particularly in the absence of specific modalities. The diffusion model excels at generating missing modalities through cross-modalities \cite{RN448}. For instance, Saharia et al. \cite{RN267} developed a multi-task diffusion-based model that addresses colorization, uncropping, and JPEG restoration, guiding image generation to ensure outputs align with specific tasks and consistently produce high-fidelity results. Furthermore, text-driven image-to-image translation \cite{RN387} utilized a pre-trained text-to-image diffusion model, generating images that correspond to text descriptions while maintaining a semantic layout from a reference image through spatial feature manipulation and self-attention. Sasaki et al. \cite{RN327} introduced a technique that learns to create images by minimizing domain-specific errors while considering the other domain's information, ensuring realistic and consistent output across both domains. Similarly, Su et al. \cite{RN388} independently trained models for source and target domains to facilitate image generation from the source to the target. Additionally, Li et al. \cite{A26} proposed the Brownian Bridge Diffusion Model (BBDM), which maps source and target images directly without relying on traditional conditional generation methods.
\subsubsubsection{Inpainting.}
Inpainting restores hidden image parts using visible data. DDPMs have been also used to solve image inpainting problems. Effective inpainting relies on transferring information from visible to hidden pixels, creating coherent and contextually consistent results. To improve the consistency of the inpainting region, Zhang et al. \cite{RN474} proposed a method called CoPAINT. This approach applies a Denoising Diffusion Implicit Model (DDIM) with a Bayesian framework to adjust both the visible and hidden parts of the image during each step of the denoising process, resulting in a more consistent and visually accurate inpainted image. RePain \cite{RN266} addressed form inpainting \cite{RN265, RN264} by enhancing specific image regions based on an arbitrary binary mask, using a pre-trained unconditional DDPM as a generative prior. By selectively adjusting reverse diffusion iterations, it refines the generation process without directly modifying or conditioning the primary DDPM network. As a result, the model produced high-quality, diverse images suitable for inpainting applications. 
\subsubsubsection{Segmentation.}
Image segmentation plays a crucial role in image analysis, facilitating the decomposition of complex images into meaningful segments \cite{RN211, RN420, RN423}. To explore the importance of the diffusion model in image segmentation, Wang et al. \cite{RN416} introduced DFormer, a diffusion-guided transformer framework for image segmentation. DFormer combines noisy masks and deep pixel-level features, utilizing masked attention in a transformer for accurate ground-truth mask identification. In short-sequence semantic segmentation, a limited set of images is utilized to classify each pixel into predefined categories. DifFSS \cite{RN417} leverages a diffusion model to incorporate diverse auxiliary support images, which offer a range of supplementary details during image generation. This methodology enhances image segmentation, by addressing structural and unstructured imbalances, thereby leading to a more robust and discriminative representation. Gu et al. \cite{RN418} illustrated the applicability of the diffusion model in instance segmentation, utilizing instance-targeted filters and unified mask branch features to regenerate global instance masks. Moreover, Baranchuk et al. \cite{RN212} showcased the utility of the diffusion model in semantic segmentation, specifically by extracting feature maps at various scales from the decoder of U-Net and concatenating them for enhanced segmentation results. Wang et al. \cite{RN421} employed the SegRefiner method to optimize the segmentation process, improving the accuracy of the input mask through a diffusion denoising approach. The primary objective was to elevate the precision and quality of object masks generated by diverse segmentation models. In a parallel approach, MaskDiff \cite{RN422} introduced a mask distribution method designed explicitly for few-short instance segmentation scenarios. This approach employs a Conditional Diffusion Probabilistic Model (CDPM) to model a binary distribution based on the RGB object region. The aim was to optimize the segmentation process and improve the quality of object masks.
\subsubsubsection{Classification.}
Incorporating the diffusion model into existing frameworks such as Contrastive Language-Image Pretraining (CLIP) and transformer is vital for image classification \cite{RN433, RN430, RN431, RN432}. Mukhopadhyay et al. \cite{RN424} utilized a diffusion model to enhance feature representation by leveraging embeddings from intermediate feature mappings, which contain crucial discriminative information for accurate classification. Similarly, Wang et al. \cite{RN425} introduced DF-GMM, an end-to-end model designed for fine-grained image classification by addressing low-rank feature maps. Moreover, Chen et al. \cite{RN426} developed the Robust Diffusion Classifier (RDC) which maximizes data likelihood and predicts class probabilities through Bayes’ theorem. They also created a multi-head diffusion backbone, modifying the final layers of the U-Net model to improve noise prediction and enhance sampling strategies. Additionally, an unsupervised framework, Diff-HSI \cite{RN427} concurrently learns spectral and spatial features for Hyperspectral Image (HSI) classification. This framework efficiently extracts spectral features across various wavelengths while capturing the spatial arrangement of objects. By employing a diffusion model for feature extraction, Diff-HSI demonstrates improved adaptability and accuracy in discerning intricate spectral-spatial relationships. Furthermore, Sigger et al. \cite{RN428} introduced DiffSpectNet which merges a diffusion spectral-spatial network with a transformer to enhance unsupervised learning, refining features using a pre-trained denoising U-Net. This combination establishes a robust paradigm for HSI, emphasizing accuracy and adaptability. Finally, Guo et al. \cite{RN429} applied the EGC energy-based model, that unifies discriminative and generative learning by leveraging probabilistic modeling and diffusion process. This model effectively estimates noise from scaled, noisy images, leading to high-quality synthesis and competitive classification accuracy.
\subsubsubsection{Editing.}
Editing images poses great challenges and is prone to distortion during the denoising process \cite{RN436, RN438, RN439}. Hou et al. \cite{RN434} improved image editing by incorporating a rectifier module that fine-tunes the diffusion model's weights. This technique enhances the model performance by compensating for missing information using residual features, which allows the network to concentrate on learning the difference between the input and the desired output. Similar to denoising score-matching, this learning approach minimizes errors during the editing process, leading to high-fidelity results in the final denoising steps. 
\subsubsubsection{Denoising.}
Image denoising enhances image quality by minimizing undesired noise and preserving crucial details for manipulation and analysis. Diffusion generative models are pivotal in the progression of denoising solutions. Xie et al. \cite{RN440} applied a diffusion model for image denoising tasks, redefining the diffusion process steps according to specific noise models. Their proposed model training strategy and sampling algorithms accommodate Gaussian, Gamma, and Poisson noise types. Wang et al. \cite{RN441} presented Reconstruction and Generation Diffusion Model (RnG) to improve sequence process fidelity. The model integrates a diffusion algorithm to amplify high-frequency details, enhancing visual quality. To manage unwanted textures and artifacts, they introduced an adaptive step control to regulate the number of inverse steps in the diffusion model.
\subsubsubsection{Discussion.}
In this section, we review several papers on the application of DDPM in natural image analysis, summarizing key findings in Fig. \ref{fig:taxnomy} and Table \ref{fig:table4}. While DDPM excels in producing high-quality and diverse outputs through iterative denoising process. Nevertheless, advanced models have emerged to reduce latency without compromising generative quality. Additionally, image translation is critical, especially when specific image modalities are difficult to obtain due to costs or constraints, such as converting day-time scenes into night-time or foggy conditions for autonomous driving training. The diffusion model also proves valuable for image inpainting, ensuring coherence in generated regions to facilitate smooth visual transitions and correct hidden patterns during denoising. Enhancing input masks through DDPM can further optimize segmentation in natural images. Moreover, integrating spectral and spatial features is crucial for accurately representing complex image structures, enhancing adaptability and accuracy in classification. Improvements in training strategies and sampling techniques can refine the quality of generated images, making the investigation of DDPM's performance and efficiency in natural image processing a promising area for future research. With recent advancements in DDPM methods, continued innovation and contributions in this field are anticipated.

\begin{table*}[htpb]
\centering
\scriptsize 
\renewcommand{\arraystretch}{1.5} 
\setlength{\tabcolsep}{4pt} 
\caption{An overview of respective studies related to DDPM in natural imaging, focusing on their datasets, methodologies, contributions, and applications.}
\begin{tabular}{ p{0.5cm} p{1.9cm} p{2cm} p{7cm} p{1cm}} 
\hline
\textbf{Year} & \textbf{Dataset} & \textbf{Method} &\textbf{Highlights} & \textbf{Application} \\
\hline
2022 & ImageNet \cite{D1}  & CDM \cite{RN228} & Conditioning augmentation to improve sample fidelity. & Generation\\
2023 & Text2Human \cite{D2} &  UPGPT \cite{RN393} & Multimodal enables generation and post editing of a person. & Generation\\
2022 & AFHQ \cite{D5, D6} & DiSS \cite{RN395} & Guidance-driven image generation from strokes and sketches. & Generation\\
2023 & Imagen \cite{D9} & Self-Guidance \cite{RN400} & Guiding diffusion models via internal representations. & Translation\\
2022 &  landscapes \cite{D11} & DiffuseIT \cite{RN448} & Self-attention for preservation and accelerated semantic style transfer. & Translation\\
2022 & ImageNet \cite{D1} & Palette \cite{RN267} & It excels in colorization, inpainting, uncropping, and JPEG restoration. & Translation\\
2021 & Facade \cite{D13} & UNIT-DDPM \cite{RN327}  & An I2I translation model generates realistic outputs from unpaired data. & Translation\\
2023 & HQ Datasetc\cite{D16} & BBDM\cite{A26} & I2I translation using Brownian Bridge bidirectional diffusion process. & Translation\\
2023 & CelebA-HQ \cite{D17} & DiffI2I \cite { RN452} & Compact prior extraction network, and transform for efficient I2I. & Inpainting\\
2023 & CelebA-HQ \cite{D17} & CoPAINT \cite{RN474} & Image inpainting by reducing mismatch using a Bayesian framework. & Inpainting\\
2023 & Texel \cite{D18} & DINAR \cite{RN265} & Neural diffusion model for realistic avatar reconstruction. & Inpainting\\
2024 &  Cityscapes \cite{D20} & DiffuMask\cite{RN420} & Cross-attention for high-resolution masks, reducing annotation costs. & Segmentation\\
2023 & RefCOCO \cite{D21} & Ref-Diff \cite{RN423} & Enhances performance, and complements discriminative models. & Segmentation\\
2023 & Fss-1000 \cite{D23} & DifFSS \cite{RN417}  & FSS using diffusion process and diverse auxiliary support images. & Segmentation\\
2022 & COCO \cite{D7} & DiffusionInst \cite{RN418} & Instance segmentation through noise-to-filter diffusion. & Segmentation\\
2024 & COCO \cite{D7}& MaskDiff \cite{RN422} & Few-Shot Instance Segmentation Model with diffusion for binary mask. & Segmentation\\
2023 & ImageNet \cite{D1} & Imagen model \cite{RN431} & Imagen model improved performance with synthetic data. & Classification\\
2024 & CIFAR-10 \cite{D27} & RDC \cite{RN426} & Generative classifier enhancing adversarial robustness via diffusion. & Classification\\
2024 & Indian Pines \cite{D31} & MTMSD \cite{RN427} & The diffusion model refines multi-timestep features for HSI analysis. & Classification\\
2023 & ImageNet \cite{D1} & EGC \cite{RN429} & The unified model combines energy-based generation and classification  &Classification\\
2023 & COCO \cite{D7} & InstructEdit \cite{RN436} & Generates segmentation prompts, edited via Stable Diffusion. & Editing\\
2023 & COCO \cite{D7} & UCE \cite{RN439} & Closed-form solutions enable scalable edits in conditional models. &  Editing \\
2023 & DIV2K \cite{D8} & Denoising \cite{RN440}  & Generative image denoising through diffusion for posterior estimation. & Denoising\\
2023 & DIV2K \cite{D8} & RnG \cite{RN441}& Denoising network with adaptive diffusion for visual enhancement. & Denoising\\
2022 & ImageNet \cite{D1} & DRL \cite{RN471} & Score matching objective for better detail control in representations. & Denoising\\
\hline
\end{tabular}
\medskip
\label{fig:table4}
\end{table*}
\subsubsection{\textbf{DDPMs for Medical Image Analysis.}}
DDPMs have garnered considerable attention in medical image analysis. High-quality medical images are crucial for accurate diagnosis and effective prognosis, directly influencing healthcare professionals' ability to detect, monitor, and treat diseases.

\subsubsubsection{Generation.}
The main goal of DDPM is image generation, particularly for synthetic histopathology images used in classifying and analyzing tissue. Probabilistic models have been used to enhance the generation of high-quality brain cancer histopathology images. Color normalization aids in learning morphological patterns and emphasizing structural details during diffusion processing \cite{RN298}. Synthetic images are vital for privacy and augmenting medical imaging data. Diffusion probabilistic models outperform GANs in generating high-quality 3D medical data, especially in MRI and CT imaging. Khader et al. \cite{RN445} explored Medical Diffusion to improve the generation capabilities of 3D data, enabling synthetic images that enhance segmentation models, even with limited training data. This makes diffusion models promising for AI development with synthetic medical data. Kim et al. \cite{RN297} focused on a diffusion and deformation module that captures spatial transformations between images, ensuring accurate alignment and improving 4D cardiac MRI sequence generation. In medical imaging, human organs continuously change due to factors such as heartbeat and aging. Generative models often neglect sequential dependencies and face challenges with high dimensionality. Yoon et al. \cite{RN396} addressed this by introducing sequence-aware deep generative models that incorporate image sequences during inference. 
\subsubsubsection{Translation.}
Obtaining multi-modal images is essential for accurate diagnosis in medical imaging. Generative models play a key role in cross-modal image translation, especially between commonly used modalities like CT and MRI. Although CT is widely employed, its limitations in detecting soft tissue injuries often necessitate MRI scans for a more thorough diagnosis. Ozbey et al. \cite {RN450, RN451, RN452} introduced the SynDiff model, employing an adversarial diffusion approach to improve medical image translation, specifically emphasizing enhancing Multi-contrast MRI to MRI-CT translation. SynDiff combines elements of GAN and Diffusion models, offering a promising solution to address current imaging limitations and advance the field of medical image translation. Li et al. \cite{RN449} proposed the Frequency Decouple Diffusion Model (FDDM) for unsupervised MR-CT image translation. By separating the frequency components in the Fourier domain, FDDM ensures realistic CT images while preserving anatomical structures. They also utilized the Frequency Guided Diffusion Model (FGDM) \cite{RN389} to enhance structure preservation during image translation by leveraging both low and high-frequency information. FGDM was trained on target-domain images, which enables zero-shot image translation by effectively transferring knowledge from the target-domain to the source-domain without requiring additional training on source-domain images. This allows for seamless cross-domain translation when applied directly to images from the source domain. The integration of DDPM \cite{RN266, RN208} and score-based diffusion model \cite{RN209} were introduced by Lyu et al. \cite{RN271} to address  MRI to CT image translation challenge. It demonstrates the superior effectiveness of the diffusion model and score-matching approach. The study \cite{RN389} addresses a major limitation in diffusion models: the loss of critical structural details during image processing. To counter this, the researchers proposed a frequency-guided diffusion model that employs specialized filters, which effectively preserve image quality. These filters excel in handling the intermediate details, where most differences between source and target images arise while maintaining the basic and fine details.
\subsubsubsection{Reconstruction.}
Efficient medical image reconstruction plays a crucial role in accurate image analysis. Numerous medical imaging techniques encounter challenges such as low-quality images, the absence of crucial anatomical details, and slow generation. Lengthy MRI, Positron Emission Tomography (PET), and CT scan time cause patient care delays \cite{RN453}. Xie et al. \cite{RN281} presented MC-DDPM, a novel approach for under-sample MRI reconstruction. This method uses a diffusion process with a conditioned mask and provides uncertainty estimates. It highlights superior performance. MC-DDPM excels in accelerated MRI reconstruction and shows potential for other under-sampled medical image tasks. Addressing the need for faster MRI, Korkmaz et al. \cite{RN455} proposed a novel technique called self-supervised Diffusion Reconstruction (SSDiffRecon). This method employed a conditional diffusion model incorporating a cross-attention transformer block \cite{RN404}. This cross-attention block focuses on a specific part of the data for effective denoising and utilizes data consistency projection to ensure that the denoised output remains aligned with the underlying structure of the input data. Additionally, the self-supervision strategy allows training using only under-sampled data, enhancing efficiency while preserving data integrity. Peng et al. \cite{RN286} introduced DiffuseRecon for MR image reconstruction, which leverages observed signals and a pre-trained diffusion model. It eliminates the need for additional training specific to the acceleration factor. Its stochastic nature allows users to visualize a range of possible reconstructions, providing diverse and representative results. 
\subsubsubsection{Segmentation.}
In medical imaging, image segmentation plays a vital role by dividing complex images into distinct regions or segments, facilitating the identification of abnormalities \cite{RN404}. Advancements in technology have led to the exploration of diffusion models to improve segmentation outcomes. Bozorgpour et al. \cite{RN401} presented DermoSegDiff, a diffusion model that incorporates boundary information during the learning process. This approach aims to enhance the segmentation of skin lesions. In clinical practice, precise segmentation masks are vital for supporting radiologists. The novel technique  BerDiff \cite{RN402} significantly improves accuracy in medical image segmentation. BerDiff employs Bernoulli noise as a diffusion kernel, introducing randomness through the sampling of initial noise and latent variables to produce a range of diverse segmentation masks. This process effectively highlights the region of interest, thereby accelerating the segmentation process. To mitigate the computational demands, a memory-efficient patch-based diffusion model (PatchDDM) was introduced \cite{RN409}. This model is designed to deal with large 3D volumes, making it more suitable for medical applications. Evaluation using CT 3D images revealed that PatchDDM provides effective 3D segmentation while significantly reducing computational resource requirements compared to traditional diffusion models.
\subsubsubsection{Classification.}
Medical image classification serves as a fundamental tool for interpreting medical data and categorizing it into predefined classes. However, studies \cite{ RN411, RN415} emphasized the significance of generative models in medical image classification. Yang et al. \cite{RN270} introduced DiffMIC, which is a framework that leverages a diffusion-based model to enhance the classification accuracy of medical images by mitigating unwanted noise and perturbation inherent in low-contrast imaging modalities. The method employs a Dual-granularity Conditional Guidance (DCG) model to capture high-level local and global features. Furthermore, the approach integrates condition-specific Maximum-Means Discrepancy (MMD) regularization to enhance classification performance. In a related study, Oh et al. \cite{RN412} presented the DiffMIX, which is based on the diffusion method designed to enhance the segmentation and classification of nuclei within pathology images characterized by significant imbalances. The method employs a diffusion model for data synthesis, utilizing two virtual patch types to augment data distribution. Subsequently, a semantic-label-conditioned diffusion model generates realistic samples, demonstrating superior performance in classification and segmentation. Shen et al. \cite{RN413} explored integrating a conditional diffusion model with a coupled transformation to enhance robustness and reliability in medical image classification. The approach involves three stages: Shallow Mapping for consistent representation learning, Corrective Diffusion to refine preliminary predictions, and a novel nested ensemble technique that combines a transformer, which learns hierarchical invariant features, with a diffusion model, which estimates distributions conditioned on these features.  This comprehensive methodology aims to address key challenges and improve performance in medical image classification.
\begin{table}
  \caption{An overview of the reviewed DDPM in medical imaging, focusing on their datasets, methodologies, contributions, and modalities.}
  \label{tab:freq}
  \centering
  \scriptsize
  \begin{tabular}{c c c c c c}
    \toprule
    Year & Dataset & Input & Method & Highlights & Modality \\
    \midrule
    2022 & TCGA \cite{D36} & 2D & Diffuse Model \cite{RN298} & Generative models for high-quality brain cancer image synthesis. & Histology \\
    2023 & TCIA \cite{D37} & 3D & Medical Diff. \cite{RN445} & Synthetic images improve AI privacy, augment datasets, realism. & MRI \\
    2023 & ACDC \cite{D39} & 3D & SADM \cite{RN396} & Sequence-aware transformer in diffusion for images. & MRI \\
    2023 & Chest X-ray \cite{D40} & 2D & GH-DDM \cite{RN392} & Leverages DDPM transform to learn the spatial relationship. & X-ray \\
    2023 & Gold Atlas \cite{D43} & 3D & DDMM-Synth \cite{RN451} & MRI-guided diffusion model integrates CT measurements. & MRI \\
    2023 & SynthRAD \cite{D44} & 3D & FDMM \cite{RN449} & Guide a diffusion model using MR information for CT generation. & MRI \\
    2023 & Lung Dataset \cite{D46} & 3D & FGDM \cite{RN389} & Structure-preserving translation using frequency-domain filters. & MRI \\
    2022 & fastMRI \cite{D47} & 3D & MC-DDPM \cite{RN281} & Accelerated MRI reconstruction with uncertainty quantification. & MRI \\
    2024 & fastMRI \cite{D47} & 3D & SSDiffRecon \cite{RN455} & SSL undersampled k-space improves MRI reconstruction. & MRI \\
    2022 & fastMRI \cite{D47} & 3D & DiffuseRecon \cite{RN286} & Coarse-to-fine Monte-Carlo sampling scheme for reconstruction. & MRI \\
    2023 & Biobank-card. \cite{D49} & 2D,3D & DMCVR \cite{RN453} & Generative model enhances resolution of cardiac image synthesis. & cMRI \\
    2023 & ISIC 2018 \cite{D50} & 2D & DermoSegDiff \cite{RN401} & Skin lesion segmentation emphasizing boundary information. & Dermoscopy \\
    2023 & BRATS 2021 \cite{D42} & 3D & BerDiff \cite{RN402} & Bernoulli noise in diffusion model improves segmentation. & MRI \\
    2024 & BraTS2020 \cite{D55} & 3D & PatchDDM \cite{RN409} & Memory-efficient patch-based diffusion reduces generation cost. & MRI \\
    2023 & APTOS2019 \cite{D56} & 2D & DiffMIC \cite{RN270} & Dual conditional guidance improves semantic representation. & OCT \\
    2023 & GLySAC \cite{D57} & 2D & DiffMIX \cite{RN412} & Generate virtual patches for nuclei class diversity and quality. & Histology \\
    2023 & ACDC \cite{D59} & 3D & FSDiffReg \cite{RN459} & Unsupervised image registration uses diffusion for deformation. & MRI \\
    2021 & ACDC \cite{D59} & 3D & DiffuseMorphs \cite{RN269} & Enables synthetic image registration with score scaling. & MRI \\
    2023 & LDCT-PD \cite{D61} & 2D & DiffMIC \cite{RN270} & Cascaded diffusion model to generate high-quality CT images. & CT \\
    2024 & Prostate-Bio. \cite{D62} & 3D & Fast-DDPM \cite{RN464} & Poisson flow and consistency denoising improves low-dose CT. & MRI \\
    2022 & OCT \cite{RN293} & 3D & OCT b-scan \cite{RN293} & Unsupervised diffusion model denoises self-fused OCT B-scans. & OCT \\
    2024 & BraTs \cite{D42} & 3D & MedSegDiff \cite{L13} & Dynamic Conditional Encoding enhances image segmentation. & MRI \\
    \bottomrule
  \end{tabular}
  \label{fig:table_DDPM_Medical_1}
\end{table}

\subsubsubsection{Registration.}
In medical imaging, precise image registration is essential for merging data from diverse scans (MRI, CT, X-rays, PET, ultrasound) and aligning images acquired at different times. DDPM provides remarkable advancement in medical image registration. Qin et al. \cite{RN459} introduced FSDiffReg, an innovative unsupervised method for non-rigid image registration. It employs Feature-wise diffusion-guided (FDG) and Score-wise diffusion-guided (SGD) modules for effective guidance and superior topology preservation for medical cardiac image registration tasks. In addition, Kim et al. \cite{RN269} introduced a novel diffusion-based approach called DiffuseMorphs by integrating DDPM with a transformation network based on U-Net. In this framework, the diffusion network evaluates the variations between moving and static images, while the transformation network uses this information to estimate the transformation field.
\subsubsubsection{Denoising.}
In medical imaging, the presence of noise and artifacts can significantly compromise the quality of the image. These noisy images potentially lead to the loss of essential diagnostic information. As a result, accurate interpretation and diagnosis become challenging in healthcare. To address these issues, diffusion generative models are a versatile solution for a range of denoising problems \cite{RN460, RN292}. Liu et al. \cite{RN463} proposed a novel unsupervised denoising method for low-dose CT images, exclusively employing normal-dose CT images to achieve noise reduction without training data. This approach effectively addresses challenges related to data scarcity in clinical settings by leveraging a cascaded unconditional diffusion model and adaptive coefficient adjustment.
Additionally, \cite{RN464} explained posterior sampling poisson flow consistency models (PS-PFCM) denoising techniques for CT. This method involves training PFGM++ \cite{RN465} to learn a noise-to-target trajectory and distillation, which transfer knowledge into PFCM for efficient CT image denoising. Hu et al. \cite{RN293} explored an unsupervised diffusion probabilistic model to denoise retinal optical coherence tomography (OCT) B-scans (2D cross-sectional image). The process involves applying Gaussian noise to self-fused, combining multiple OCT B-scans, and utilizing a Markov chain for adjustable denoising levels.
\subsubsubsection{Discussion.}
This subsection explored the application of DDPM in medical image analysis. DDPMs are prominent in generation tasks, enhancing medical imaging by capturing finer details, as illustrated in  Table \ref{fig:table_DDPM_Medical_1}. They facilitate stable, reliable images, improving clinical decision-making in complex situations involving noise reduction, image reconstruction, and anatomical structure preservation. Additionally, DDPMs play a vital role in image translation across various modalities, such as CT, MRI, ultrasound, and X-rays, ensuring that critical details remain intact for accurate diagnosis and treatment planning. By producing consistent, denoised outputs, DDPMs enhance the quality of medical image segmentation and classification. For example, the DiffMIX \cite{RN412} method addresses challenges in segmenting and classifying nuclei in pathology images. Although noise can lead to potential information loss, DDPMs can still generate high-quality, realistic images.
\subsection{LDM for Natural and Medical Image}
LDMs are widely adopted in both natural and medical image applications, as they enable faster processing through compressed latent representations and dimensionality reduction.
\subsubsection{Generation.}
The impressive capability of LDMs to balance generative quality with efficiency makes them valuable for various applications. By leveraging the latent space of a pre-trained autoencoder, they improve visual fidelity while optimizing computational resources, demonstrating competitive performance in various generation tasks \cite{RN199}. UMM-Diffusion \cite{L5} explored an image generation method that uses both text and image together, combining them into a single multi-modal latent space for improved guidance. Corneanu et al. \cite{RN264} introduced LatentPaint to enhance the computational speed of DDPM by using a forward-backward fusion step in latent space, reducing data dimensionality, training costs, and improving inference speed. Moreover, Pinaya et al. \cite{RN294} through exploring LDMs \cite{RN199} for synthesizing images from high-resolution 3D brain scans, capturing complex structures for high-fidelity medical image synthesis. Additionally, CoLA-Diff \cite{RN447}, a multi-modality MRI synthesis model leverages conditional LDM and deviates from the traditional single-modality approaches by introducing a memory-efficient multi-modal framework. This model incorporates morphological guidance during the diffusion process to preserve anatomical configurations and enhance synthesis quality while improving model efficiency.
\subsubsection{Translation.}
Multimodality plays a crucial role across numerous applications, particularly in transforming one type of image into another, such as converting sketches to realistic photos. To address this, Ding et al. \cite{L6} introduced latent optimization techniques to refine noisy latent features, aligning reference images with generated outputs for controlled image synthesis. This method enables sketch-to-image generation and real-image editing guided by sketches, without requiring additional training or fine-tuning. LaDiffGAN \cite{L7} addressed unsupervised image-to-image (I2I) translation by integrating a GAN in the diffusion latent space, enhancing inference speed. Moreover, MS-SPADE \cite{RN450} utilized Latent Diffusion Models (LDM) to translate a single-source modality into multiple target modalities in 3D medical imaging. It achieved this by transforming the source latent into the target latent through style transfer.
\subsubsection{Reconstruction.}
Image reconstruction aims to recover or enhance the quality of the image from degraded inputs. He et al. \cite{RN453} innovatively reconstructed 3D cardiac volumes using a morphology-guided diffusion model, eliminating iterative optimization and enhancing generation quality. The learned latent spaces offer interpretable global and local cardiac information for precise 3D shape reconstruction from 2D cMRI slices.
\subsubsection{Segmentation.}
In addressing the challenges associated with handling large volumes of unlabelled data in medical imaging, a new diffusion-based approach utilizing a multi-level consistency model (MGCC) has been proposed \cite{RN403}. The approach leverages the LDM \cite{RN199} to generate images, incorporating varying levels of global context noise perturbation applied to the input of an auxiliary decoder to assist in the decoding process. By maintaining consistency between multiple decoders, the method ensures output stability and enhances representational capabilities. The MGCC model has demonstrated significant improvements in segmentation accuracy by effectively transferring distribution knowledge. In a related study, Kim et al.\cite{RN407} employed a self-supervised diffusion adversarial representation learning (C-DARL) for blood vessel segmentation and vascular disease diagnosis. This model was trained on synthetic vessel images generated from the diffusion latent technique and achieved performance comparable to supervised methods. Despite its efficiency, the model's requirements for substantial 3D high-resolution data posed significant computational challenges, limiting its applicability in practical medical scenarios. 
\subsubsection{Classification.}
LDM reduces images to a lower-dimensional latent space, which enables models to focus on the most discriminative elements for classification.
 Besides, Packhauser et al. \cite{RN446} introduced a privacy-preserving sampling technique designed to exclude synthetic images when a particular patient profile has been recognized from the source training data. In addition, they utilized an LDM \cite{RN199} in conjunction with this privacy-enhancing sampling strategy to generate X-ray images. These images were then employed in a thoracic abnormality classification task within chest radiography. To enhance colorectal cancer classification performance in histopathology, Niehues et al. \cite{L8} applied LDM with varying parameter configurations, selecting the optimal settings to generate an augmented dataset for histopathology classification. Abstreiter et al. \cite{RN471} introduced a Diffusion-Based Representation Learning (DLR) solution involving a denoising score matching leverage trainable encoder, ensuring meaningful representation in the latent space without extensive modifications for better classification.
\subsubsection{Editing.}
Furthermore, LDM facilitates local image edits through text guidance and addresses LDM reconstruction limitations by exploring thin masks for localized edits \cite{RN334}.  Expanding on text guidance for local image editing, DiffEdit \cite{RN437} employs text-conditioned diffusion models for semantic image editing, automatically generating masks based on contrast predictions from the text input’s diffusion model. Latent inference is applied to preserve content in identified regions, showcasing strong synergy with mask-based diffusion. Avrahami et al. \cite{RN335} presented a method that enables precise image editing through natural language descriptions. This method enables users to define specific areas for modification using a mask, while Contrastive Language-Image Pretraining (CLIP) offers guidance to create images based on text descriptions. After the denoising process, the mask is applied to the latent images, integrating noise from the original images to ensure the final output accurately reflects the intended changes. Svitov et al. \cite {RN265} explored DINAR, a novel framework for generating rigged full-body avatars from single RGB images. They leveraged the LDM \cite{RN199, RN205}, demonstrating its effectiveness in training within the neural texture space. Neural textures are learnable image-based representations, that enhance the visual realism of 3D objects or avatars by encoding complex surface details, such as fine texture, lighting, and shading effects. This approach enables realistic reconstruction of extensive unseen regions, such as a person's back, starting from a frontal view. The models in their pipeline were adapted to train using 2D images and videos.
\subsubsection{Discussion.}
LDMs are considered powerful tools in various image analysis tasks. Remarkably, LDM has demonstrated superior performance compared to DDPM and WDM in solving trilema problem \cite{RN238}, which includes fast sampling, high-quality samples, and diversity. Nevertheless, the application of LDM is often constrained by several challenges. One significant issue is the potential loss of finer details during the encoding process into the latent space. Additionally, the architectural complexity of LDM requires substantial computational resources, which can be a limiting factor to adopt in complex tasks. The model’s performance heavily depends on the effectiveness of the latent space representation, if the latent space fails to capture essential features adequately, it can degrade overall model efficacy. 
{\scriptsize
\begin{table}
\centering
\caption{An overview of the reviewed LDMs in natural imaging, focusing on their datasets, methodologies, contributions, and applications.}
\begin{tabular}{c c c c c}
  \toprule
  Year & Dataset & Method & Highlights & Application \\
  \midrule
  2022 & CelebA-HQ \cite{D17} & LDMs \cite{RN199} & Enhance visual fidelity using limited resources through latent space. & Generation \\
  2024 & CelebA-HQ \cite{D17} & LatentPaint \cite{RN264} & Training-free inference using latent space forward-backward fusion. & Generation \\
  2024 & Diffsound \cite{D19} & FLAN-T5 \cite{RN205} & Enhance visual fidelity using limited resources through latent space. & Generation \\
  2023 & LAION-400M \cite{D32} & UMM-Diff. \cite{L5} & Generates customized images by combining text and image inputs. & Generation \\
  2024 & Sketchy \cite{L9} & Latent Opt. \cite{L6} & Latent optimization refines noisy latents using cross-attention. & Translation \\
  2024 & U-GAT-IT \cite{L10} & LaDiffGAN \cite{L7} & Enhance GANs with latent encoding and conditional denoising loss. & Translation \\
  2022 & CIFAR10 \cite{D27} & DLR \cite{RN471} & Diffusion model encodes details denoising, enhancing classification. & Classification \\
  2021 & LAION-400M \cite{D32} & Blended Lat. \cite{RN334} & Accelerated text-driven image editing via optimized LDM reconstruction. & Editing \\
  2024 & DRAGBE \cite{RN438} & DragDiff. \cite{RN438} & Optimized latent diffusion for enhanced point-based image editing. & Editing \\
  2022 & ImageNet \cite{D1} & DiffEdit \cite{RN437} & Automated mask generation for image edits using text conditioning. & Editing \\
  2021 & LAION-400M \cite{D32} & Blended Diff. \cite{RN335} & Blend noised input with diffusion latent augmentations for better realism. & Editing \\
  \bottomrule
\end{tabular}
\label{fig:table_LDM_Natural}
\end{table}
}
{\scriptsize
\begin{table}
\centering
  \caption{An overview of the reviewed LDM in medical imaging, focusing on their datasets, methodologies, contributions, and modalities.}
  \begin{tabular}{c c c c c c}
    \toprule
    Year & Dataset & Input & Method & Highlights & Modality \\
    \midrule
    2022 & Lung Dataset \cite{D46} & 2D & CoLA-Diff \cite{RN447} & Structure-preserving zero-shot image translation using diffusion. & CT \\
    2022 & Biobank MRI \cite{D35} & 3D & Brain-LDM \cite{RN294} & Explores LDM for generating synthetic brain images from 3D MRI. & MRI \\
    2024 & IXI dataset \cite{D47} & 3D & MS-SPADE \cite{RN450} & Single model enables versatile, multi-modal MRI translation. & MRI \\
    2023 & Biobank MRI \cite{D35} & 3D & DMCVR \cite{RN453} & Morphology-guided DMCVR reconstructs 3D cardiac volume. & cMRI \\
    2024 & XCAD \cite{D54} & 2D & C-DARL \cite{RN407} & C-DARL generates synthetic blood vessel images using contrastive. & X-ray \\
    2022 & ChestX-ray \cite{D40} & 2D & $\mathrm{Syn}_{\mathrm{LDM}}$ \cite{RN446} & Latent diffusion model synthesizes privacy-enhanced chest X-rays. & X-ray \\
    2023 & BRATS2018 \cite{D41} & 3D & Adaptive LDM \cite{RN450} & Switchable block for style transfer 3D medical images. & MRI \\
    2023 & BUSI Dataset \cite{D52} & 2D & MGCC \cite{RN403} & Improved semi-supervised segmentation using LDM samples. & US \\
    2022 & Gold Atlas \cite{D43} & 3D & Diff-Score \cite{RN297} & Models generates frames along continuous path using latent space. & MRI \\
    2024 & TCGA-BRCA \cite{D78} & 2D & PathLDM \cite{L15} & Text-conditioned LDM for histopathology image generation. & Histology \\
    \bottomrule
  \end{tabular}
   \label{fig:table_LDM_Medical}
\end{table}
}
LDM holds considerable promise for enhancing image generation tasks, it is essential to address the drawbacks for optimal performance.Future research directions could focus on developing strategies to overcome these challenges, thereby unlocking the full potential of LDM in practical applications. Overall, LDM has more positive potential, while there is a downside that also exists and is carefully considered

\subsection{WDM for Natural and Medical Image}
The ability of WDM to capture multiscale wavelet features makes it highly valuable in medical imaging, where precise details are essential for diagnosis. Additionally, WDM is being explored to reduce latency for both natural and medical image analysis.
\subsubsection{Generation.}
Diffusion models are crucial for high-fidelity image generation, yet training and inference times hinder real-time applications. To tackle these challenges, Phung et al. \cite{RN233} combined wavelet transforms with diffusion models, effectively reducing the dimensionality of wavelet subbands to accelerate the diffusion process without compromising generative quality. By decomposing images into four subbands, the model can learn from the wavelet spectrum, leading to faster and more stable training. While wavelet transformations are typically associated with image compression, as in JPEG-2000 \cite{RN246}, they also highlight diffusion models' limitations, such as DDPMs' bias against high-frequency generation, affecting accurate frequency representation. Spectral Diffusion (SD) \cite{RN262} addressed this by employing wavelet gating to enhance high-frequency generation and improve texture detail, facilitating low-cost image generation. Hui et al. \cite{RN259} further advanced this approach by merging diffusion with wavelet coefficients for rough 3D images, using a compact wavelet technique to sample from random noise for unique shapes. Zhou et al. \cite{A31} introduced UDIFF, which is a 3D diffusion model that leverages unsigned distance fields (UDFs) and wavelet transformations to efficiently generate textured 3D shapes in the spatial-frequency domain. In a parallel effort, Guth et al. \cite{RN256} introduced the Wavelet Score-based Generative Model (WSGM), enhancing generative processes by leveraging conditional probabilities of multiscale wavelet coefficients. Similarly, Gal et al. \cite{RN247} combined Style-GAN with wavelet transformations, reducing computational complexity and improving spectral fidelity in high-resolution models. Building on this, Ding et al. \cite{A30} developed Counter Wavelet Diffusion, which decomposes images into low- and high-frequency components and employs an attention module to enhance high-frequency details, preserving image quality while reducing processing time through the capture of both positional and frequency information. Addressing the challenges of noise and data scarcity in medical imaging, Wu et al. \cite{A29} proposed a wavelet-enhanced denoising method within a score-based generative model (SGM), which improved noise robustness and training stability for reconstructing undersampled and noisy data. Furthermore, Friedrich et al. \cite{A27} advanced 3D wavelet-based diffusion for medical image synthesis by integrating wavelet coefficients into the diffusion process, enabling efficient generation of high-resolution 3D medical images.
\begin{figure}[t]
    \centering\includegraphics[width=1\textwidth]{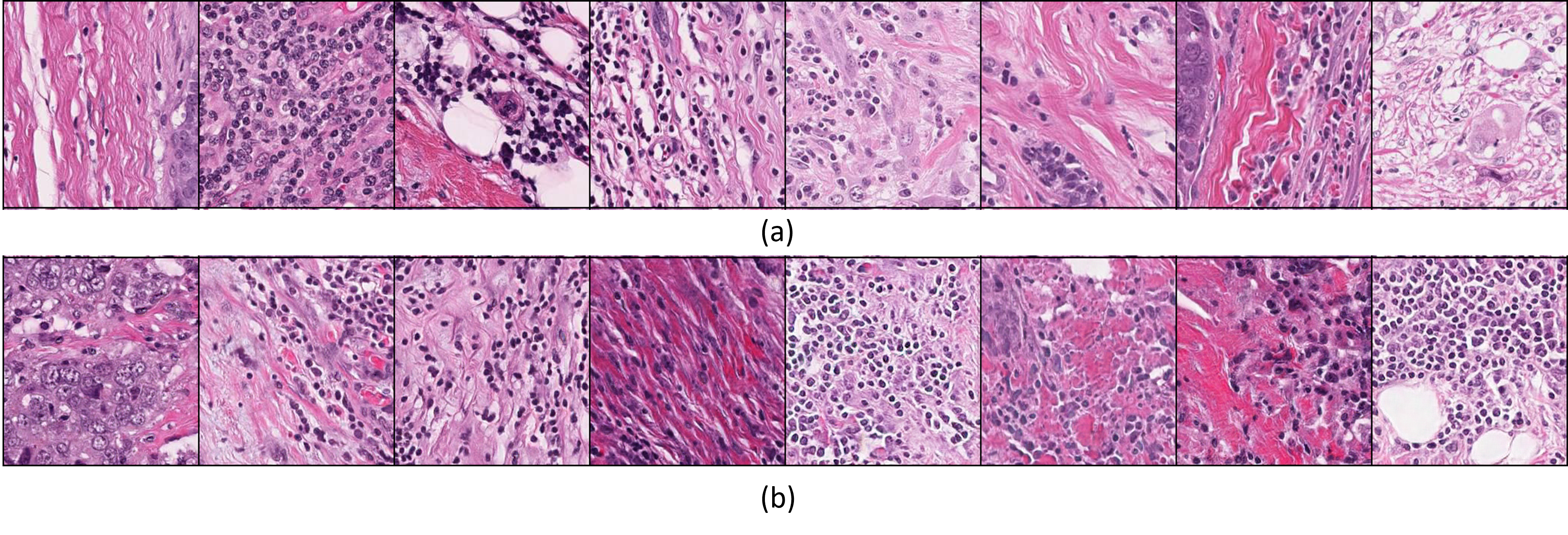}
    \caption{Illustrate histopathology images generated using WDM: (a) shows a real image, and (b) presents a synthetic image.}
    \label{fig:WSGM}
    \Description{A real and generated histopathology image.}
\end{figure}
\subsubsection{Reconstruction.}
Learning feature distribution in the wavelet frequency domain notably reduces computational costs compared to the spatial domain. Introducing a wavelet-diffusion, and a unique training strategy \cite{RN233}, accelerates training by fourfold, ensuring efficient restoration. The Wavelet domain allows both low and high-frequency components to aid in learning distinct features, enhancing the model's capability for effectively reconstructing images. Addressing this, Jiang et al. \cite{RN257} introduced Diffusion-based Low-Light (DiffLL) based on a distinctive Wavelet-based Conditional Diffusion Model (WCDM). This model harnesses the generative capabilities of diffusion models and the strengths of Wavelet transform for robust and efficient image enhancement. Notably, it incorporates a unique training strategy ensuring content consistency during inference by engaging both forward and denoising training phases. Recognizing the significance of high-frequency components, the model adeptly captures diagonal details essential for effective reconstruction.
WaveDM, developed by Huang et al. \cite{RN258}, offers an efficient image reconstruction solution by leveraging wavelet diffusion and conditional sampling to minimize sampling steps. It learns the distribution of clean images in the wavelet domain, conditioning the Wavelet spectrum of degraded images. This domain choice reduces computational time and memory, outperforming traditional spatial methods. More recently, the wavelet-inspired multi-channel diffusion model (WISM) \cite{RN483} has been introduced to identify diverse image patterns across various frequency components while maintaining spatial correspondence with the original images. By integrating Wavelet domain and image domain diffusion models, this approach ensures both consistency and validation in the reconstruction process, particularly for limited-angle CT reconstruction.
\subsubsection{Translation.}
Wavelet diffusion applications in the medical domain, exemplified by studies like knowledge distillation in image-to-image translation \cite{RN255}, highlight their relevance. Wavelet transforms excel in emphasizing frequency information while reducing sampling time significantly and leveraging frequency sparsity and dimensionality reduction. They used Wavelet transforms to condense high-dimensional spaces into low-dimensional manifolds, which is crucial for faster and more efficient sampling processes. This approach enhances frequency awareness in both image and feature domains.
\subsubsection{Denoising.}
In a pioneering approach, Moser et al. \cite{RN479} introduced DiWa a diffusion wavelet approach, to address the trade-off between image quality, particularly in preserving high-frequency details and computational efficiency. This method enhances super-resolution images by synthesizing high-frequency information within the wavelet spectrum, resulting in high-quality, denoised outputs.
\subsubsection{Classification}
Li et al. \cite{RN242} focusses on noise suppression to tackle imager classification problem. This methodology integrates Wavelet transform seamlessly with CNN techniques. This approach decomposes the feature map into low and high-frequency components during downsampling. The lower-frequency component retains the main features, while the high-frequency component, containing most data noise, is discarded during inference. This strategic elimination enhances noise robustness and contributes to more effective image classification. 
Liu et al. \cite{RN243} introduced an multi-level Wavelet CNN strategy,  diminishing the resolution of the feature map. This reduction is coupled with an enlarged receptive field, ultimately enhancing the model's capabilities in image reconstruction, denoising, and overall image reconstruction tasks. The versatile application of Wavelet transform has led to its widespread use in various domains of computer vision, including image restoration, \cite{RN244}, visual representation learning \cite{RN245}, and image compression \cite{RN246}, With the remarkable advancement in visual generative tasks mainly due to the rapid progress of generative model.
The Wavelet transform emerges as a potent time-frequency transformation tool capable of describing images at multiple resolutions. This capability allows for a more comprehensive representation of intricate details in the image, making the Wavelet transform a valuable asset in addressing the challenges associated with high-frequency features during image generation \cite{RN250}.
{\scriptsize
\begin{table}
\scriptsize
\caption{An overview of the reviewed WDM in natural imaging, focusing on their datasets, methodologies, contributions, and applications.}
\centering
\begin{tabular}{c c c c c}
\toprule
\textbf{Year} & \textbf{Dataset} & \textbf{Method} &\textbf{Highlights} & \textbf{Application} \\
\midrule
2023 & CelebA-HQ \cite{D17} & WDM \cite{RN233} & WDM enhances generation and convergence through adaptive processing. & Generation\\
2022 & CIFAR-10 \cite{D27} & SD \cite{RN262} & Spectral Diffusion uses wavelet gating for frequency features  recovery. & Generation\\
2022 & CelebA-HQ \cite{D17} & SPI-GAN \cite{RN476} & Enhanced GAN model accelerate through straight-path interpolation. & Generation\\
2024 & DeepFash.\cite{D65} & UDiFF \cite{A31} & 3D diffusion model with optimal wavelet transform for texture generation. & Generation\\
2022 & CelebA-HQ \cite{D17} & WSGM \cite{RN256} & Accelerates SGMs by factorizing data into conditional wavelet coefficients. & Generation\\
2021 & FFHQ \cite{D11}  & SWAGAN \cite{RN247} &  Hierarchical WDM enhances high-frequency generation efficiently. & Generation \\
2024 & CIFAR-10 \cite{D27}  & CWDM \cite{A30} & CWDM enhances facial expression generation efficiency. &Generation\\
2023 & LOLv1 \cite{D67} & DiffLL \cite{RN257}  & Model combines diffusion and wavelet transforms for stable generation. & Reconstruction\\
2023 & RainDrop \cite{D68}, & WaveDM \cite{A28} & ECS models clean image, and optimizing sampling steps. & Reconstruction\\
2021 & CelebA-HQ \cite{D17}  & WaveFill \cite{RN186} & Wavelet-based inpainting uses DWT to resolve frequency conflicts. & Inpanting \\
2024 & GoPro \cite{D77}  & HWDM \cite{L12} & Restores frequency information using multi-scale extraction. & Denoising \\
2024 & FFHQ \cite{D11}  & DiWa \cite{RN479} & Combining diffusionand wavelet transforms for superior image quality. & Denoising \\
\bottomrule
\end{tabular}
\label{fig:table_WDM_natural}
\end{table}
}

{\scriptsize
\begin{table}
  \caption{An overview of the reviewed WDM in medical imaging, focusing on their datasets, methodologies, contributions, and modalities.}
  \label{tab:table_WDM_medical}
  \centering
  \begin{tabular}{c c c c c c}
    \toprule
    Year & Dataset & Input & Method & Highlights & Modality \\
    \midrule
    2023 & Kvasir \cite{D76} & 2D & LLCaps \cite{RN487} & Multi-scale design preserves resolution, wavelet attention enhances features. & Endoscopy \\
    2023 & ADNI \cite{D38} & 3D & W-Diffusion \cite{L11} & Enhances data availability using DDPM and wavelet decomposition. & MRI \\
    2023 & fastMRI \cite{D47} & 3D & W-SGM \cite{A29} & Enhances reconstruction stability and accuracy with wavelet-SGM. & MRI \\
    2024 & BraTs \cite{D42} & 3D & 3D-WDM \cite{A27} & Efficient 3D WDM generates high-resolution images effectively. & MRI \\
    2024 & fastMRI \cite{D47} & 3D & WaveDM \cite{A28} & Learns clean images in the wavelet domain for efficient restoration. & MRI \\
    2024 & AAPM \cite{D75} & 2D & SWORD \cite{RN482} & Unified model optimizes wavelet frequency components for training. & CT \\
    2024 & AAPM \cite{D75} & 2D & WISM \cite{RN483} & Wavelet transform enhances SGM for improved LA-CT reconstructions. & CT \\
    \bottomrule
  \end{tabular}
  \label{fig:table_WDM_medical}
\end{table}
}


\subsubsection{Discussion:}
WDM are emerging in various fields, as shown in Fig. \ref{fig:taxnomy}. Recent trends and advancements can be observed in Table \ref{fig:table_WDM_natural}, which highlights the significant potential of WDM in natural images, and Table \ref{fig:table_WDM_medical}, which focuses on medical images. WDM achieves dimensionality reduction by a factor of four, effectively quadrupling processing speed. Operating in both spectral and spatial domains, it facilitates multiscale feature extraction while preserving the structural integrity of images during the diffusion process. This preservation minimizes the loss of critical texture information, which is essential in medical imaging. Consequently, WDM supports the generation of high-quality synthetic images, enhances resolution, and reduces noise, thereby aiding accurate diagnoses. In practical applications like CT and MRI, 3D WDM enhances image reconstruction from incomplete or noisy datasets, effectively reducing artifacts and improving contrast. Similarly, WDM is beneficial for tasks in natural images, such as super-resolution and deblurring, by capturing details like edges while mitigating noise. Additionally, when paired with diffusion models, wavelet transforms improve image quality during decompression, suggesting a promising future for WDM applications in medical imaging.
\section{General Discussion}
This paper presents a comprehensive overview of DDPM, LDM, and WDM, examining their efficiency and generative quality with various real-world applications. We identified both the strengths and limitations of these models, particularly in the context of medical imaging applications. As shown in Table \ref{fig:table1} and Fig. \ref{fig:taxnomy}, our survey paper uniquely addresses model adoption in varied scenarios within both natural and medical imaging. Additionally, we discussed the applicability and adoption of these models across various scenarios in natural and medical imaging. Our findings suggest that DDPM, for instance, is more suited for tasks where high-quality generation is critical, but processing time is less of a concern. Conversely, LDM proves effective in situations where both quality and processing time are important. When the preservation of fine details and texture with speed is prioritized, WDM is the most suitable model. In light of these findings, it is clear that model selection should be application-specific, as each model's performance and relevance vary depending on the task. Nonetheless, our analysis reveals considerable potential for further refinement and advancement within each diffusion model.

Moreover, medical image diagnosis is a complex process that relies on advanced computational techniques. The analysis of medical images involves generating high-quality visuals, which requires sophisticated algorithms, such as diffusion models shown in Fig. \ref{fig:taxnomy}.  The challenge lies in handling large volumes of data, improving image clarity, and accurately detecting abnormalities. This intricate nature of medical imaging and the complexity of the algorithms pose significant challenges for researchers. Moreover, medical imaging data is often limited, making it challenging to obtain labeled datasets for training deep learning algorithms \cite{A43}. These models rely on large datasets for effective training, where diffusion-based generative models (DDPMs, LDM, and WDMS) come into play; they help increase the amount of available data to address this shortfall. In the medical field, these models are significant due to concerns about data privacy and security \cite{A44}. Handling sensitive patient information raises privacy issues, especially when this data is shared between institutions or stored on external servers for training. The diffusion model addresses these challenges by generating high-quality images without compromising patient privacy. However, creating high-resolution images that maintain critical medical details, such as texture and contrast is complex, especially with pathological images. The WDM \cite{RN233} effectively tackles this problem, ensuring the generated images retain the necessary medical intricacies.

As we observed in Table \ref{table:3}, natural image generation has reached a more advanced stage, with most state-of-the-art models evaluated on well-established, widely-used datasets. Natural images are generally less costly and readily accessible, making them easier to obtain. In contrast, medical images pose unique challenges; they are more expensive to acquire due to dependencies on specialized annotations, privacy constraints, and the involvement of medical professionals. This complexity makes generating high-quality, efficient synthetic medical images significantly challenging, and researchers are still striving to enhance the generative quality in this field. The following section outlines current challenges and potential solutions, presenting these issues as open challenges for future research.

\section{ Challenges and Future Prospects }

Although these models bring many benefits, there are associated challenges that limited their performance. In the following, we will summarize the challenges and discuss about potential solutions. Then we will explore future research opportunities.
\subsection{Challanges.}
\subsubsection{Slow sampling speed.}
The diffusion model faces a significant challenge in slow sampling speed due to its iterative process, which involves a series of stepwise denoising stages. This process makes the model computationally intensive and time-consuming, limiting its practicality for applications that require fast generation, such as real-time systems or deployment on devices with limited resources.
\subsubsection{High dimensionality.} While powerful in image generation, diffusion models pose limitations in high-dimensional medical images such as 3D or WSI. These limitations include slower generation processes due to iterative inference steps, limited applicability to certain data types, and the inability to reduce dimensionality. However, these constraints do not diminish the unique strengths of diffusion models in generating high-quality images without labeled data. However, these challenges catalyze future research to optimize inference times and improve model utility. The advancements in diffusion models promise prospects and challenges in computationally efficient medical imaging.
\subsubsection{Balancing speed and quality.}
While GANs excel in rapidly producing realistic samples, their mode convergence is limited. Conversely, VAEs provide diversity but often need better sampling quality. The emergence of diffusion models addresses the shortcomings of both VAEs and GANs, offering adequate mode convergence and high-quality sampling. However, their iterative nature results in a slow sampling process, making them practically expensive and requiring further refinement \cite{RN260}.
The ongoing challenge lies in balancing generation speed and image quality. Future research should prioritize exploring innovative methodologies or hybrid models that optimize both aspects simultaneously. Discussed approaches, such as LDM offer the potential to achieve a balance between generative speed and output quality, but structural preservation is challenging, incorporating wavelet transforms addresses this by enhancing structural fidelity. The main aim is to achieve swift generation while preserving the high-quality imaging capabilities inherent in diffusion models.
\subsubsection{Faster converges and more stability.}
The traditional diffusion model faces challenges in practical training due to its reliance on a mix of coarse and detailed information. Conversely, the Wavelet diffusion \cite{RN233} approach addresses this issue by partitioning input images into frequency bands. This technique provides multiscale information and can capture distinctive features by modeling stable training. Additionally, achieving generalization across various datasets and tasks, despite differences in image characteristics and noise levels, further complicates the process.
\subsubsection{Lack of interpreatbility.}
Lack of interpretability in the diffusion model can be a significant drawback, especially for application in medical imaging, where understanding the decision-making process of a model is crucial. Diffusion models, with their intricate training dynamics and high-dimensional data generation, can often operate as black-box systems, making it difficult for medical professionals to understand how the model arrives at its conclusions. This lack of transparency raises concerns about trust and accountability, particularly when these models are used for critical tasks such as diagnosing diseases or recommending treatment plans. Improving the interpretability of these models is essential for their adoption in high-stakes fields like healthcare, where decisions can have life-altering consequences.
\subsection{Future outlook.}
\subsubsection{Improve sampling efficiency.}
Improving sampling in diffusion models is a promising future prospect. Efforts to enhance sampling efficiency and quality will make diffusion models more practical for real-world applications, particularly in fields like medical imaging where faster, accurate output is crucial. Approaches like wavelet-based regularization, adaptive sampling, and combining diffusion with GAN are potential directions to achieve more efficient and high-quality sampling. 
\subsubsection{Enhancing efficiency in medical imaging.}
Future research should focus on further improving the efficiency of diffusion models, especially in the context of medical imaging. Using efficient diffusion models like LDM \cite{RN199} and Wavelet diffusion \cite{RN233} proves promising, enhancing speed without compromising generative quality. This efficiency is crucial for applications such as the generation of pathology WSIs, as presented by Harb et al. \cite{RN468}. The computational complexity arising from high resolution and large image size poses a significant challenge. Consequently, the integration of Wavelet diffusion stands out as a prospective approach to expedite the generative process of WSIs in the future.
\subsubsection{Representation space and features.}
Diffusion models need help to generate semantically meaningful data representations within their latent space, thereby limiting their utility for tasks that demand semantic understanding \cite{RN193, RN470}. This issue arises from information loss during the diffusion process. To overcome this limitation, models capable of learning semantically rich representations are necessary, as they contribute to enhancing image reconstructions and semantic interpolations. The diffusion-based representation learning \cite{RN471} concept is shaped by an information-theoretic encoder that modulates the input information passed to the score function throughout the diffusion process. Meanwhile, Jiang et al. \cite{RN257} introduce a lightweight image enhancement approach using wavelet-based conditional diffusion, emphasizing high-frequency restoration for improved fine-detail reconstruction. Preserving crucial information in medical imaging at a minor level during the diffusion process is crucial. Nevertheless, there needs to be a suitable representation in the latent space for wavelet diffusion for researchers.
\section{Conclusion.}
In this paper, we comprehensively reviewed the literature concerning an efficient diffusion model from various perspectives. We initiated with an examination of the foundational aspect of DDPMs \cite{RN266}, encompassing their current developments and challenges. The survey highlights the rapid growth of diffusion models within medical imaging, demonstrating superior image fidelity and sampling speed performance. Our investigation extends to the general framework of DDPMs, LDMs \cite{RN199} and WDMs \cite{RN233} methods. Following this, we summarized the strengths and weaknesses inherent to each model, providing a comprehensive comparative analysis. Furthermore, we provided a critical analysis of each model’s contributions across various fields of application, exploring multiple application directions and shed light on how these efficient and high-quality generative models enhance real-world applications. 
We also present a comprehensive comparative analysis between our survey paper and existing surveys shown in Table \ref{fig:table1} and Fig. \ref{fig:taxnomy}, which provides a broad picture for the research community. The section on prospects and challenges delineates promising avenues for future research in natural and medical imaging.


\bibliographystyle{unsrt}

\bibliography{sample-base}

\end{document}